\documentclass[manuscript,authorversion,nonacm]{acmart}

\AtBeginDocument{%
  }

\setcopyright{rightsretained}
\copyrightyear{2024}
\acmYear{2024}
\acmDOI{XXXXXXX.XXXXXXX}

\acmPrice{15.00}
\acmISBN{978-1-4503-XXXX-X/18/06}

\usepackage{tcolorbox}
\usepackage{enumerate}
\usepackage{eurosym}
\usepackage{booktabs}
\usepackage{tcolorbox}
\tcbuselibrary{hooks, breakable, skins}
\usepackage{enumitem}
\usepackage{subcaption}
\usepackage{longtable}

\newcommand{\note}[1]{
	\begin{tcolorbox}[
        width=\linewidth,
        breakable, enhanced,
		leftrule=3pt, toprule=1pt, rightrule=1pt, bottomrule=1pt, 
		arc=0mm, nobeforeafter,  colback=white]	
		#1
	\end{tcolorbox}
}

\def\dbar#1{
  {\color{gray}\rule{#1mm}{6pt}}} 
\def\ybar#1{
  {\color{teal}\rule{#1mm}{6pt}}} 
\def\nbar#1{
  {\color{purple}\rule{#1mm}{6pt}}} 
\def\hbar#1{\hspace{-0.6mm}{\color{gray}\rule{1.2mm}{#1pt}}}

\usepackage{tabularx}
\usepackage{xspace}

\usepackage[capitalise]{cleveref} 

\newcommand{\numberofparticipantsOne}{$57$\xspace}
\newcommand{\numberofparticipantsTwo}{$47$\xspace}

\newcommand{\accessed}{\textit{last accessed October 15, 2024}}

\begin{document}


\title[Assessing Perception and Usage of Biometrics]{Do They Understand What They Are Using? -- Assessing Perception and Usage of Biometrics}


\author{Lukas Mecke}
\affiliation{%
 \institution{University of the Bundeswehr Munich}
  \city{Munich}
  \country{Germany}}
  \email{lukas.mecke@unibw.de}
\affiliation{%
 \institution{LMU Munich}
  \city{Munich}
  \country{Germany}}
  \email{lukas.mecke@ifi.lmu.de}

\author{Alia Saad}
\affiliation{%
 \institution{University of Duisburg-Essen}
  \city{Essen}
  \country{Germany}}
\email{alia.saad@uni-due.de}

\author{Sarah Prange}
\affiliation{%
 \institution{University of the Bundeswehr Munich}
  \city{Munich}
  \country{Germany}}
  \email{sarah.prange@unibw.de}

\author{Uwe Gruenefeld}
\affiliation{%
 \institution{University of Duisburg-Essen}
  \city{Essen}
  \country{Germany}}
  \email{uwe.gruenefeld@uni-due.de}

  \author{Stefan Schneegass}
\affiliation{%
 \institution{University of Duisburg-Essen}
  \city{Essen}
  \country{Germany}}
  \email{stefan.schneegass@uni-due.de}

\author{Florian Alt}
\affiliation{%
 \institution{University of the Bundeswehr Munich}
  \city{Munich}
  \country{Germany}}
  \email{florian.alt@unibw.de}

\renewcommand{\shortauthors}{Mecke et al.}

\begin{abstract}

In this paper we assess how well users know biometric authentication methods, how they perceive them, and if they have misconceptions about them. We present the results of an online survey that we conducted in two rounds (2019, N=57; and 2023, N=47) to understand the impact of the increasing availability of biometrics on their use and perception. The survey covered participants' general understanding of physiological and behavioral biometrics and their perceived usability and security. While most participants were able to name examples and stated that they use biometrics in their daily lives, they still had difficulties explaining the concepts behind them. We shed light on participants' misconceptions, their coping strategies with authentication failures and potential attacks, as well as their perception of the usability and security of biometrics in general. As such, our results can support the design of both further studies to gain deeper insights and future biometric interfaces to foster the informed use of biometrics.

\end{abstract}

\ccsdesc[500]{Security and privacy~Usability in security and privacy}
\ccsdesc[300]{Security and privacy~Biometrics}

\keywords{physiological biometrics, behavioral biometrics, authentication, usable security}

\maketitle

\section{Introduction}
Biometric authentication methods provide rich opportunities for fast, seamless, and secure authentication on many devices that are omnipresent in users' daily lives. Examples include fingerprint~\cite{jain1999multichannel} and face recognition~\cite{vazquez2016face}, available on most modern smartphones (for example, ``Face ID'' on iPhones), or gait recognition~\cite{sprager2015inertial} as part of Google's ``Smart Lock'' feature on Android\footnote{\url{https://support.google.com/android/answer/9075927}, \accessed}.
%
At the same time, biometric systems are inherently based on machine learning models, making them prone to external factors~\cite{bhagavatula2015biometric} and biases~\cite{drozdowski2020demographic} as well as intransparent and hard to predict~\cite{wojciech2017explainable}.

While experts and designers may be aware of those points, this knowledge does not necessarily translate to end-users~\cite{adams1999, asgharpour2007}. 
As some examples, it is important to understand that physiological biometrics cannot be changed, making the system potentially insecure should they be leaked. Behavioral biometrics on the other hand are less permanent and often require retraining after some time to remain usable and secure. Without such knowledge, users can form wrong mental models leading to insecure behavior~\cite{wash2010soups}. They may lose trust when encountering such problems, be unable to access their devices, or -- in the worst case -- be harmed~\cite{yapo2018ethical}. 

Other fields of security have already shown how better understanding their users can help to improve the interaction with authentication and shape the design of user interfaces. Collecting commonly used passwords leads to password policies and approaches to aid users in understanding the strength of their passwords (e.g. \cite{seitz2017pasdjo}). Similarly, knowledge of how users chose graphical patterns allowed for approaches to nudge users to a more secure choice (e.g. \cite{bulling2012increasing}).
This paper aims to create a similar knowledge base to support and inform the design of biometric interfaces. 

However, the perception of biometrics may also evolve with the availability and adoption of new technology~\cite{franks2021changing}, so sampling a single point in time may not be sufficient. To address this, we propose to compare the perception and use of biometrics for two points in time. 
To this end, we designed and conducted an online survey in two rounds (first round, 2019: $N = \numberofparticipantsOne$, second round, 2023: $N = \numberofparticipantsTwo$) assessing participants' knowledge about biometrics, their current use of this technology, and their perception of security and usability. We explicitly covered both, physiological and behavioral biometrics to be able to make a comparison. 

We found that most participants indicated being unable to define or explain biometrics. However, our open-text answers revealed that many participants were able to name correct examples and have some basic understanding while in-depth knowledge was often lacking. In the second round of our survey, more participants actively used biometrics for authentication. 
However, contrary to our expectations we did not observe other clear effects between the rounds. Behavioral biometrics seemed overall less known in both rounds and were also consistently rated worse than physiological mechanisms. 
We discuss those results and give practical implications; in particular, suggesting ways to foster user literacy and control of biometric methods.

\textbf{Contribution Statement.}
We contribute 1) an online survey conducted in two rounds assessing users' perception and use of biometric methods. We 2) identify common themes and misconceptions, and 3) discuss how our insights can be used to improve biometric interfaces and foster future informed use of biometric methods.

\section{Background and Research Approach}
In this section, we introduce the concept of biometrics and some forms they appear in as well as give an overview of some previous work on understanding their perception. We conclude by highlighting the research gap and how our work contributes to closing it. 

\subsection{Types of Biometrics}
The term biometrics refers to mechanisms to recognize individuals (for example, for authentication) based on distinctive biological traits~\cite{jain1996introduction,jain2004introduction,jain2010biometrics}. These are broadly divided into \emph{physiological} or the physical traits of a person, and \emph{behavioral} traits, focusing on the specific way in which a person moves or interacts with their surroundings~\cite{sharif2019overview,dargan2020comprehensive, liebers2020introducing}. 
There is a plethora of work introducing physiological biometrics with approaches including among others face recognition~\cite{zhao2003face,tolba2006face,taskiran2020face},  fingerprints~\cite{maltoni2009handbook,marasco2014survey,yang2019security}, hand or palm prints~\cite{kong2009survey,zhong2019decade}, human iris features~\cite{de2016iris,bharathi2020review} or footprints~\cite{wang2017single, hotfoot}. Behavioral biometrics are receiving increasing attention as well, most notably the recognition of individuals based on their walking (i.e., gait)~\cite{gafurov2007survey,wan2018survey,singh2018vision,sepas2022deep}, typing patterns (i.e., keystroke dynamics)~\cite{buschek2015, karnan2011biometric,teh2013survey,raul2020comprehensive}, or touch dynamics~\cite{teh2016survey,ellavarason2020touch}.

\subsection{Understanding Biometrics Perception}
Large parts of the literature about biometrics focus on technical aspects like data collection (e.g., ~\cite{phillips2017lessons,stylios2022biogames}), feature extraction (e.g.~\cite{galar2015survey}) or classification (e.g.~\cite{yampolskiy2008behavioural,sundararajan2018deep}). Here we give an overview of work investigating user perception. 

Furnell and Evangelatos~\cite{furnell2007public} used Likert scale questions to understand users' awareness and usage of biometrics, finding that participants preferred methods they had previously heard of and considered easy to use. In a study by Elliott et al.~\cite{elliott2007perception} participants voiced their concerns when using biometrics including cleanliness of the devices, safety, and which applications and who would have access to their data.  
Bhagavatula et al.~\cite{bhagavatula2015biometric} compared the usability of fingerprint, face recognition, and PIN under different conditions. They found fingerprint to be the overall preferred method with mixed results for face recognition (e.g. because it was unusable in dark environments). A survey on the perception of facial recognition~\cite{ada2019beyond} found that despite 90\%  of the participants being familiar with the technology, only 5\%  claimed adequate knowledge to build a solid opinion on its usage and its implications. In a large-scale survey (N= 10,000), Franks et al.~\cite{franks2020identity} found that 76\% of the respondents used biometric technology, mainly fingerprint and facial recognition. 
Saad et al.~\cite{saad2022mask} investigated the impact of the Covid-19 pandemic on device usage and authentication in an online survey. They found that the pandemic countermeasures (e.g., sensitization measures, wearing masks) negatively affected biometric-based authentication approaches such as fingerprint and face recognition.

In a usability questionnaire on both physiological and behavioral methods, Alhussain et al.~\cite{alhussain2013users} showed that 87.3\% of participants believed that biometrics (particularly fingerprints) help to protect critical information on their phones. 
Karatzouni et al.~\cite{karatzouni2007perceptions} conducted a focus group and found that participants showed interest in adopting biometrics to enhance privacy while also having concerns about constantly being recorded. In an online survey by Rasnayaka et al.~\cite{rasnayaka2018wants}, security awareness levels reflected users' willingness to adopt biometric-dependent continuous authentication and Buckley et al.~\cite{buckley2019language} found that context is fundamental with regard to acceptability, despite the general observation that users find familiar biometrics most convenient. Sieger et al.~\cite{sieger2010poster} found that voice or speaker recognition is not suitable in crowded places.
Similarly, a survey by Ellavarason et al.~\cite{ellavarason2020touch} showed that users were concerned with external factors that might affect identification performance (for example, surrounding noise on voice recognition). In general, the fingerprint was often chosen as the most secure biometric identification approach~\cite{bhagavatula2015biometric, buckley2019language}.

\subsection{Research Gap and Summary}

Related work often focused on technical aspects of biometrics (e.g., to improve the underlying models), specific methods (e.g., only face recognition\cite{ada2019beyond}), or specific contexts and scenarios \cite{bhagavatula2015biometric,saad2022mask}. 
Furthermore, related work often used closed questions -- leaving participants less room to express their own experiences -- and assessed knowledge of biometrics after giving a definition \cite{byun2013exploring} instead of exploring their initial association.

This paper extends previous work by taking a more holistic approach and including both behavioral and physiological biometrics without a specific scenario. We allow for more open (and unprimed) responses, particularly when assessing knowledge, and investigate use in daily life and changes over time. To our knowledge, no previous work compared their results over a larger time span to observe changes.

\subsection{Research Questions}\label{sec:research_questions} 

We aim to assess which biometrics participants know and use and if they understand them. As a second step, we try to gain insights into users' perceptions of the usability and security of biometrics. Our overarching aim is to uncover potential misconceptions users of biometrics might have. This may serve to inform the design of biometric interfaces that support users in understanding biometrics and the implications of their actions as well as allowing them to act on this knowledge. 
The following research questions guide our work:

\begin{itemize}
\setlength{\itemindent}{0.5em}
    \item [RQ1] \textbf{Literacy}: Do participants know what biometrics are and can they explain how they work?
    \item [RQ2] \textbf{Perception \& Usage}: What is participants' personal view about biometrics and where do they see their value, both in their daily life and in general?
    \item [RQ3] \textbf{Usability \& Security}: How do participants perceive usability and security aspects of biometric methods and where do they see the potential for improvement?
\end{itemize}
\pagebreak
\section{Survey}

\begin{figure}
    \centering
    \includegraphics[width=\linewidth]{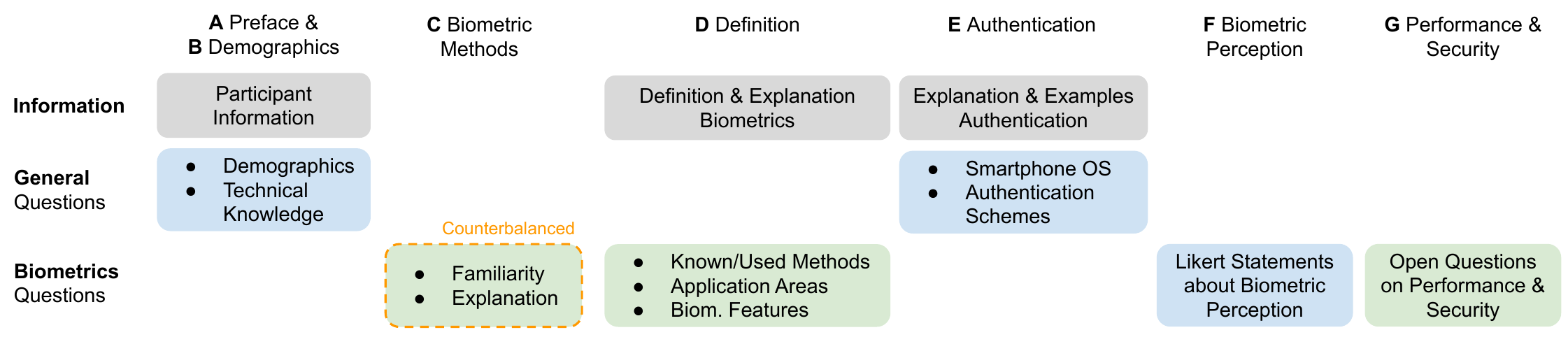}
    \caption{Visual overview of our survey structure. Participants received information and had to answer general- and biometric-specific questions. Biometric questions were repeated, starting with physiological biometrics (except for part C where they were counterbalanced). Part H is omitted for the sake of brevity. Open questions are marked green, and closed questions are marked blue. }
    \label{fig:survey_structure}
\end{figure}

Here, we introduce our study design, the survey structure, our recruitment strategy, and the data analysis approach. 

\subsection{Study Design}

Our study design follows a mixed-methods approach with two independent variables. To extend previous work, we distinguish between \textsc{types} of biometrics and explore our research questions for both \textit{physiological} and \textit{behavioral} methods. To uncover potential changes over the past years we compare quantitative results between \textsc{rounds} and employ a two-step coding approach (see Section \ref{sec:data_analysis}).

\subsection{Survey Structure}\label{sec:survey_structure}

To address our research questions (\cref{sec:research_questions}), we designed an online survey that we repeated in two rounds to explore changes over time due to the increasing prevalence of biometric methods. 
The survey comprised eight parts as described below. Refer to Appendix~\ref{app:questionnaire} for the full list of questions and to Figure \ref{fig:survey_structure} for a visual overview.

\begin{enumerate}[label=\Alph*]
    \item \textbf{Preface}: Participants were informed about the study and consented to the data collection. 
    To test awareness and understanding, it was important to ensure that participants did not look up terms or return to change their answers once they were given a definition. We asked participants to follow those guidelines and disabled returning to previous questions in the survey. 
    \item \textbf{Demographics}: This part included questions about the participant's age, gender, and occupation. Participants were also asked to estimate their technical knowledge on a 5-point Likert scale. 
    \item \textbf{Biometric Methods}: In this part, we asked participants if they were familiar with the concept of biometric methods and -- in case they were -- to explain the concept and how it works. If they were unfamiliar with the concept, we asked them to answer with their thoughts instead. 
    \item \textbf{Definition}: At this point, we gave participants the following definition of biometrics and a short explanation\footnote{Note that we did not expect participants' answers in part C to exactly match this definition, but rather wanted to ensure a common ground of knowledge for the rest of the survey.}:
    
    \note{
    \small
        \textbf{Biometrics}: 
         \textit{automated recognition of individuals based on their biological and behavioral characteristics} (ISO 2382-37)~\cite{iso2382-37}.
        In other words, a biometric system uses unique characteristics in human physiology or behavior to accurately identify individuals.
    }
    We then asked participants for biometric methods they knew, methods they used in their daily lives, and other application areas of biometrics they knew or could think of. 
    Finally, we asked them to think of other characteristics that could be used for biometric identification and give an application if they had one in mind. 
    \item \textbf{Authentication}: In this part, we explained the term authentication and gave examples for biometrics:

    \note{
    \small
        One of the main application areas for biometrics nowadays is \textbf{authentication}. That means that a user can verify their identity to, for example, access an account or device. 
        
        \noindent Common examples for \textbf{physiological} biometrics include fingerprint recognition, face recognition and iris scans.  Common examples for \textbf{behavioral} biometrics include gait recognition (walking patterns), keystroke dynamics (typing behavior), and interaction behavior (e.g. credit card usage surveillance to prevent fraud).   
    }     
    We then asked for the participants' smartphone OS and authentication scheme. 
    We moved this from the demographic questions in part B to avoid bias toward answers on smartphone authentication in C and D.
    \item \textbf{Biometric Perception}: In this part, we asked participants to answer perception questions about using biometrics on a 5-point Likert scale with the option to indicate that they did not know. 
    \item \textbf{Performance \& Security}: Here, we asked for participants' confidence in biometrics, the impact changes in their appearance or behavior might have, and their fallback strategy if the biometric system did not work. We further asked for the hack's consequences for them, their approach to attacking a biometric system (to explore their understanding of possible attack vectors), and their ideas to improve biometrics. 
    \item \textbf{Conclusion}: Finally, we gave participants the option to comment on their previous answers, leave a general comment, and leave their email to participate in the raffle. 
\end{enumerate}

We asked about both biometric methods (physiological and behavioral biometrics) in parts C, D, F, and G. In Part C, we counterbalanced the appearance of the biometric methods to avoid bias in the awareness and understanding questions based on previously answering the questions for the other biometric group. In the later part of the survey, we always started with questions for physiological biometrics and continued with behavioral biometrics. We omitted the balancing here, as definitions and examples for both groups were given in parts D and E, respectively.

Overall, we used the order of questions to avoid biases where possible and used counterbalancing where this was not the case. On the other hand, we wanted to ensure that all participants were on the same page regarding the terms used. Thus, we structured the survey to provide definitions and additional information only after we had asked for previous experiences and knowledge and before other questions where the knowledge was relevant/needed.

\begin{table}[!tb]
    \centering
    
    \setlength{\tabcolsep}{3pt}
    \begin{tabular}{lrrrrl}
    \toprule
        &\multicolumn{2}{c}{\textbf{Round 1}}& \multicolumn{2}{c}{\textbf{Round 2}}&\\
        \midrule
        \textbf{Gender}         
        & 33 &(58\%)& 26 &(55\%)   & Female \\
        & 24 &(42\%)&  15 &(32\%)   & Male\\
        & 0 & (0\%)& 6 &(13\%) & Not stated \\
        \textbf{Age}            & 29.2 &&  27.4 &   & Mean \\
                                & 18-66  &  &18-64   &  & Range\\
        \textbf{Tech. knowledge} &
        3.3 
        &\hbar{0.350877193}
        \hbar{2.105263158}
        \hbar{2.807017544}
        \hbar{3.50877193}
        \hbar{1.228070175}  & 
        3.3  &
        \hbar{0.2222222222}
        \hbar{2.222222222}
        \hbar{2.666666667}
        \hbar{4.222222222}
        \hbar{0.6666666667}  & Mean \\                            
        \bottomrule
    \end{tabular}

    \caption{Demographics of the participants of the first (N=57) and second (N=47) round of our survey. Technical knowledge was self-reported and collected on a 5-point Likert scale.}    \label{tab:demographics}
    
\end{table}

\subsection{Participants \& Recruitment}\label{sec:participants}
In both rounds, we followed the same recruiting strategy and advertised the study via social networks and university mailing lists.
We recruited two independent samples of N=\numberofparticipantsOne participants in the first round (December 2019) and N=\numberofparticipantsTwo in the second round (January 2023).
We kept the invitation (title: ``Survey about biometric perception'') and introduction to the survey on an abstract level to not prime participants prior to survey participation. We decided on independent samples for the same reason. 
The survey was conducted in English and participants in each round could voluntarily participate in a raffle for three online shopping vouchers for 30 euros each.
In both samples, participants were mostly students from non-technical fields with a slight bias towards female participants. They had a mean age between 27 (Round 2) and 29 years (Round 1) and self-reported medium technical knowledge. Table~\ref{tab:demographics} provides an overview. 

\subsection{Ethical Considerations}
In the country where this research was conducted, there is no requirement for IRB approval for this type of research.
However, we complied with all our institution's guidelines and national data protection regulations.
In particular, consent was gathered on the first page of the survey before any data was collected. Data was collected anonymously using random identifiers and stored on university servers.
Participation in the raffle of vouchers was voluntary. Email addresses were only collected for this purpose and deleted afterward.

\subsection{Data Analysis}\label{sec:data_analysis}
Four authors analyzed the responses to the open questions.
We started with the responses given in the first round, following the inductive theme generation approach for thematic analysis by Braun et al.~\cite{braun2021can}.
After an initial phase of familiarization with the provided statements, we independently applied open coding to the statements of the first round\footnote{In particular, we did not pre-assume participants' statements to exactly match the ISO definition given in \cref{sec:survey_structure} (part D), but followed an open coding approach based on the collected answers.}. 
In a review meeting, we discussed and iteratively refined the codes.
We then constructed an online affinity diagram \cite{harboe2015realworld} of these open codes
and organized them into groups, which were in a next step further refined into themes using an online whiteboard.

As a result of the analysis, we derived a codebook containing the first round's themes, groups, and codes. The same authors continued the analysis with a deductive approach by independently applying the codebook to the statements of the second survey round. Our rationale behind this two-step approach was to find differences between the rounds based on codes disappearing or new codes emerging in the second round. 
We reviewed the coding in a final meeting (see Appendix~\ref{app:codebook} for the final codebook) and resolved disagreements through discussion.
Due to the exploratory nature of our study, we refrain from reporting inter-rater agreement scores~\cite{McDonald2019reliability}.  

\subsection{Limitations}

We used two independent samples which may have induced underlying differences in the groups that did not result from the temporal distance. To minimize this effect, we exactly replicated our recruitment strategy and compared the demographics from both rounds; finding them to be very similar (see Section \ref{sec:participants}). 
Our sample was self-selected and biased towards young female students and thus, our results may not apply to the general population.
Also, security behaviors as stated in our survey might differ from participants' real-world behavior.
Lastly, experimenter bias may have had an impact on our results. To address this, four researchers were involved in the analysis of open-ended questions (see \cref{sec:data_analysis}). As such, we believe that this would not influence the resulting discussion and practical implications.

\section{Results}
We present our quantitative and qualitative findings. We structure this section based on the themes identified in our thematic analysis (see Appendix \ref{app:codebook} for the codebook) and add quantitative results from our survey where they thematically fit. We indicate the number of participants mentioning specific themes to provide a descriptive overview of our data. However, we cannot assume that participants not mentioning a specific aspect is equivalent to them not knowing the answer. Thus, we only conduct statistical tests on our quantitative results. 
We cite participants from both samples with their IDs as assigned by our survey tool and indicate the respective survey round (e.g., P12$_{R1}$ would refer to participant 12, who was part of the first round of our survey). When distinguishing between types of biometrics, we use PB and BB in subscript for physiological and behavioral biometrics, respectively. 

\begin{table*}[!tb]
    \centering
    
    \begin{tabular}{l rr@{}lrr |rr@{}lrr |r}
    \toprule
        &\multicolumn{5}{c}{\textbf{Round 1}}& \multicolumn{5}{c}{\textbf{Round 2}}&\\
        &Yes&&&No&NA &Yes&&&No&NA &\\
        \midrule
        
        \textbf{Familiar} 
        &12 (21\%)&\ybar{2.1}&\nbar{7.9}&45 (79\%)&--
        &7 (15\%)&\ybar{1.5}&\nbar{8.5}&40 (85\%)&--&PB\\
        &4 ( \ 7\%)&\ybar{0.7}&\nbar{9.3}&53 (93\%)&--
        &5 (11\%)&\ybar{1.1}&\nbar{8.9}&42 (89\%)&-- &BB\\
        
        \textbf{Access} 
        &32 (56\%)&\ybar{5.6}&\nbar{1.9}&11 (19\%)&14 (24\%)
        &30 (64\%)&\ybar{6.4}&\nbar{2.3}&11 (23\%)&6 (13\%) &PB\\
        &27 (47\%)&\ybar{4.7}&\nbar{1.1}&6 (11\%)&24 (42\%)
        &24 (51\%)&\ybar{5.1}&\nbar{0.6}&3 ( \ 6\%)&20 (43\%) &BB\\

        \textbf{Change} 
        &32 (56\%)&\ybar{5.6}&\nbar{1.8}&10 (18\%)&15 (26\%)
        &31 (66\%)&\ybar{6.6}&\nbar{1.7}&8 (17\%)&8 (17\%) &PB\\
        &32 (56\%)&\ybar{5.6}&\nbar{0.9}&5 ( \ 9\%)&20 (35\%)
        &31 (66\%)&\ybar{6.6}&\nbar{0.4}&2 ( \ 4\%)&14 (30\%) &BB\\
        
        \bottomrule
    \end{tabular}
    \caption{Participants' answers to the questions if they were \textbf{familiar} with the concept of physiological/behavioral biometrics (PB/BB), if they believed someone could \textbf{access} their device if it was protected by PB/BB, and if \textbf{changes} in their physiology/behavior would impact a PB/BB system.}
    \label{tab:quantatives}
    
\end{table*}

\subsection{Definition and Function of Biometrics}
We asked participants to define and explain, with their existing knowledge, what biometrics are and how such methods work. Participants were encouraged to guess if they were unfamiliar with biometrics.
We prefaced this open question with a binary choice, where only a minority of participants in both rounds indicated they were familiar with either physiological (familiar$_{R1}$ = 21\%,  familiar$_{R2}$ = 15\%) or behavioral (familiar$_{R1}$ = 7\%, familiar$_{R2}$ = 11\%) biometrics (Table \ref{tab:quantatives}). This constrasts the results by Buckley et al.~\cite{buckley2019language} who found 74\% of responants indicating to be familiar with biometrics and might be due to our more fine grained method of asking for physiological or behavioral biometrics instead of familiarity with biometrics more broadly. Using Fisher's exact test, we found no effects of the type of biometrics or the round on reported familiarity. 

While we did not expect participants to exactly replicate a technical definition (such as given in \cref{sec:survey_structure}, part D), we saw that many of them mentioned features related to biometrics, either remaining rather generic (e.g. naming just ``physiological features'') or giving concrete examples like face geometry or fingerprints ($N_{R1}=45$ and $N_{R2}=19$).
Some participants also explicitly mentioned where they had heard of biometrics (P181$_{R1}$: ``\textit{I only know the word biometric from ID suitable photographs}'', P98$_{R1}$:``\textit{Like in Mission: Impossible – Rogue Nation where they measure how you walk}''). Fewer participants included a correct or related verb (e.g. recognize or authenticate) in their definition ($N_{R1}=32$ and $N_{R2}=15$) or only mentioned an example related to biometrics ($N_{R1}=24$ and $N_{R2}=12$).
At the same time, a large number of participants showed some missing knowledge ($N_{R1}=48$ and $N_{R2}=34$) by either explicitly stating to have no idea ($N_{R1}=29$ and $N_{R2}=4$) or referring to other concepts. Those often revolved around related topics like medicine and health (e.g. P130$_{R1}$: ``\textit{The status of ones health in numbers}''), body functions (e.g. P129$_{R1}$: ``\textit{It could be about the way we perceive the locations of our extremities}'') or influences on behavior (e.g. P124$_{R1}$: ``\textit{trying to derive how people behave from their physical features}'', P280: ``\textit{How and why we behave as we do}''). 
Almost all participants left at least one of the questions about defining and explaining biometrics empty ($N_{R1}=53$ and $N_{R2}=42$).
Finally, some participants explicitly expressed confusion concerning the terms physiological and behavioral biometrics ($N_{R1}=2$ and $N_{R2}=9$). 
Interestingly, one participant expressed that the term physiological biometric seems to be incorrect, saying ``\textit{face, fingerprints, and such for me are anatomical characteristics, not physiological}'' (P297$_{R2}$).

\begin{figure*}[!t]
    \centering
    \begin{subfigure}[c]{.49\linewidth}
        \includegraphics[width=\textwidth]{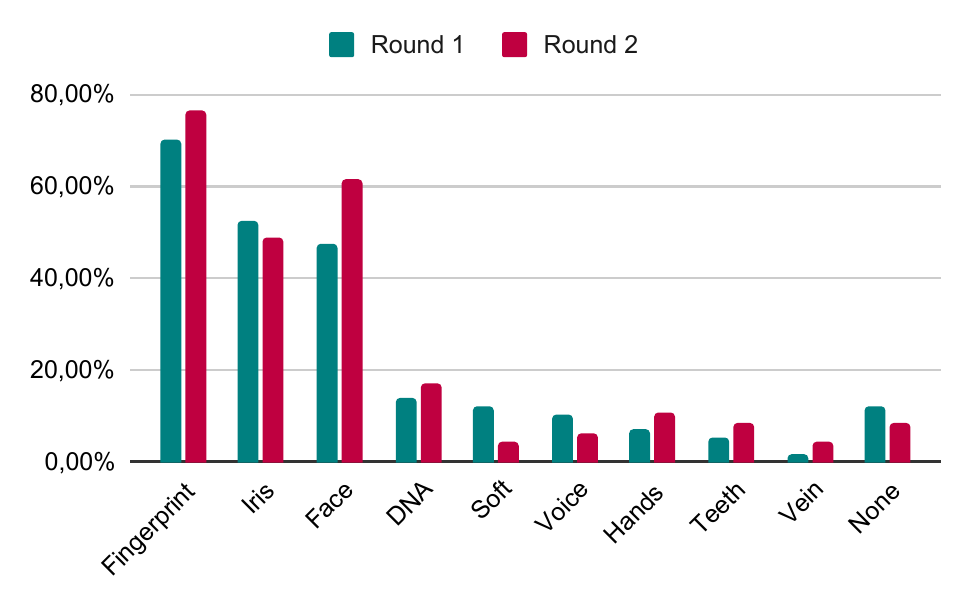}
        \subcaption{Known \textit{physiological} biometrics}
    \end{subfigure}
    \begin{subfigure}[c]{.49\linewidth}
        \includegraphics[width=\textwidth]{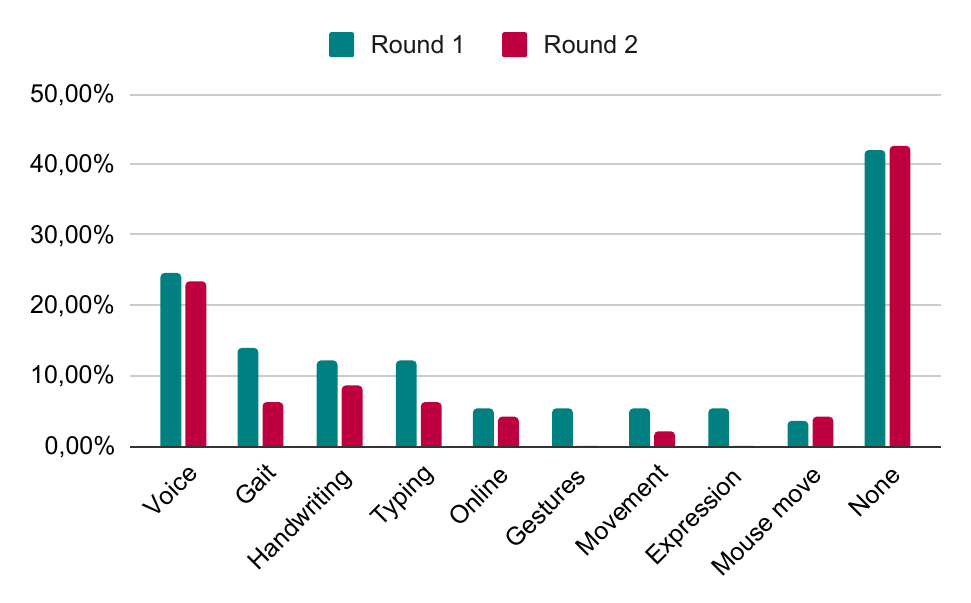}
        \subcaption{Known \textit{behavioral} biometrics}
    \end{subfigure}
    \caption{Known biometrics as mentioned by the participants. We excluded mentions of unrelated methods as well as biometrics that were mentioned by less than 3 participants across both rounds.}
    \label{fig:known}
\end{figure*}

After giving participants a definition of biometrics in the survey, we asked them to name all biometric methods they knew (see Figure \ref{fig:known}). The most mentioned physiological approaches were fingerprint ($N_{R1}=40$, $N_{R2}=36$), iris or retina scans ($N_{R1}=30$, $N_{R2}=23$), and face recognition ($N_{R1}=27$, $N_{R2}=29$). For behavioral biometric methods, the most common mention was voice recognition ($N_{R1}=14$, $N_{R2}=11$), followed by gait ($N_{R1}=8$, $N_{R2}=3$), handwriting or signatures ($N_{R1}=7$, $N_{R2}=4$), and typing or keystroke dynamics ($N_{R1}=7$, $N_{R2}=3$). 

Overall, participants knew more physiological methods, which is also reflected in the high number of participants who indicated being unable to name a single behavioral method ($N_{R1}=24$ and $N_{R2}=20$). Many participants mentioned more abstract features (e.g. gestures or online behavior) or features that are associated with biometrics but not commonly used in the consumer market. Examples are features used in a forensic context (e.g. DNA or teeth) and for profiling and tracking purposes (e.g. movement, mouse movement, online behavior). Some participants mentioned features like height or eye color (also called \textit{soft} biometrics~\cite{dantcheva2015else}) that have some biometric value but are not commonly used alone but rather in conjunction with other biometrics.
Notably, voice was mentioned both as a behavioral and physiological trait. 
While voice recognition is often considered to be a behavioral biometric method the distinction is not completely clear cut and voice does have a clear physiological component to it~\cite{jain2004introduction}.

\subsection{Perception and Usage of Biometrics}
In the survey, participants mentioned several aspects related to the usage of biometrics.
Many participants gave concrete usage examples of biometrics ($N_{R1}=29$, $N_{R2}=27$), such as fingerprint, face recognition, ID cards, and signatures, among others.
Moreover, they mentioned specific devices and use cases in which biometrics are utilized ($N_{R1}=19$, $N_{R2}=23$), primarily mentioning mobile devices and computers.
A few participants provided reasons as to why they use biometrics ($N_{R1}=4$, $N_{R2}=7$): for example, they stated that biometrics are easy, fast, safe, and less error-prone.
One participant stated to ``\textit{use the facial recognition and fingerprint scanner on [their] phone and tablet, to unlock [their] devices more easily and avoid the danger of other people seeing [their] PIN code or password}'' (P324$_{R2}$).
In contrast, a few participants stated directly that they do not use biometrics and use, for example, passwords instead ($N_{R1}=1$, $N_{R2}=3$).
Some more participants gave reasons against using biometrics ($N_{R1}=14$, $N_{R2}=7$), mentioning not having a reason for using them, having concerns about privacy and recovering from data getting compromised or pointing out issues with the recognizer/classifier. Some participants just did not like the thought of using biometrics with P20$_{R1}$ finding face recognition ``\textit{creepy and insecure}'' and P317$_{R2}$ stating to``\textit{feel more comfortable not using it}''. 

\begin{table}[!t]
    \centering

    \setlength{\tabcolsep}{3pt}
    \begin{tabular}{lrrl rrll}
    \toprule
        &\multicolumn{3}{c}{\textbf{Round 1}}& \multicolumn{3}{c}{\textbf{Round 2}}&\\
        \midrule
        \textbf{Operating}
        & 38 &(67\%)&\dbar{6.7}& 28 &(60\%) & \dbar{6}&Android\\
        \textbf{System}& 17 &(30\%)&\dbar{3}& 18 &(38\%)&\dbar{3.8} & iOS\\
        & 2 &(4\%)&\dbar{0.4} & 1 &(2\%)&\dbar{0.2} & Other\\
        \midrule
        \textbf{Unlock} 
        & 30 &(53\%)&\dbar{5.3}& 23 &(49\%)&\dbar{4.9} & Finger\\
        & 9 &(16\%)&\dbar{1.6}& 5 &(11\%)&\dbar{1.1} & PIN\\
        & 8 &(14\%)&\dbar{1.4} & 2 &(4\%)&\dbar{0.4} & Pattern\\
        & 8 &(14\%)&\dbar{1.4} & 3 & (6\%)&\dbar{0.6}& Slide/None\\
        & 2 &(4\%)&\dbar{0.4}& 2 &(4\%)&\dbar{0.4}  & Password\\          
        & 0 & (0\%)&\dbar{0}& 12 &(26\%)&\dbar{2.6} & Face\\
        \midrule
        \textbf{Fallback}
        & 28 &(51\%)&\dbar{5.1}& 25 &(53\%)&\dbar{5.1} & PIN\\
        & 15 &(26\%)&\dbar{2.6}& 7 &(15\%) &\dbar{1.5}& None\\
        & 7 &(12\%)&\dbar{1.2} & 6 & (13\%)&\dbar{1.3}& Pattern\\
        & 4 &(7\%)&\dbar{0.7} & 9 &(19\%)&\dbar{1.9} & Password\\ 
        & 2 &(4\%)&\dbar{0.4}& 0 &(0\%) &\dbar{0} & Other\\    
        \bottomrule
    \end{tabular}
    \caption{Operating system and authentication schemes used by the participants of the first and second round of our survey. }
    \label{tab:smartphone}
\end{table}

\begin{figure*}[!t]
    \centering
    \includegraphics[trim={0 25 0 0},clip,width=.48\linewidth]{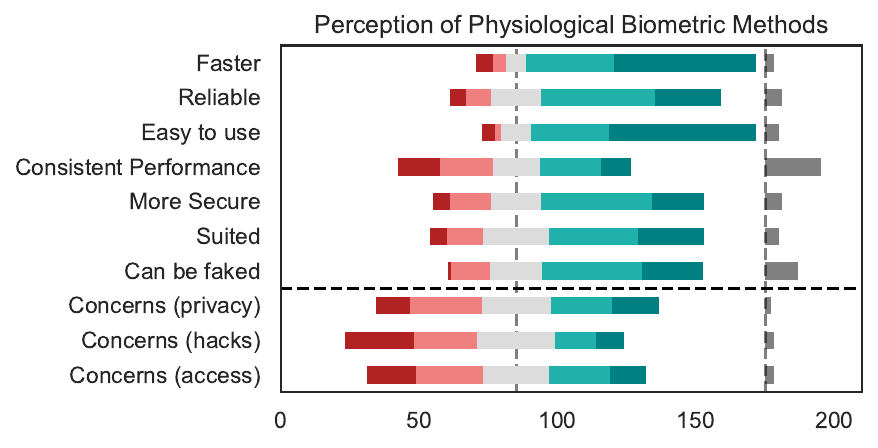}
    \includegraphics[trim={0 25 0 0},clip,width=.48\linewidth]{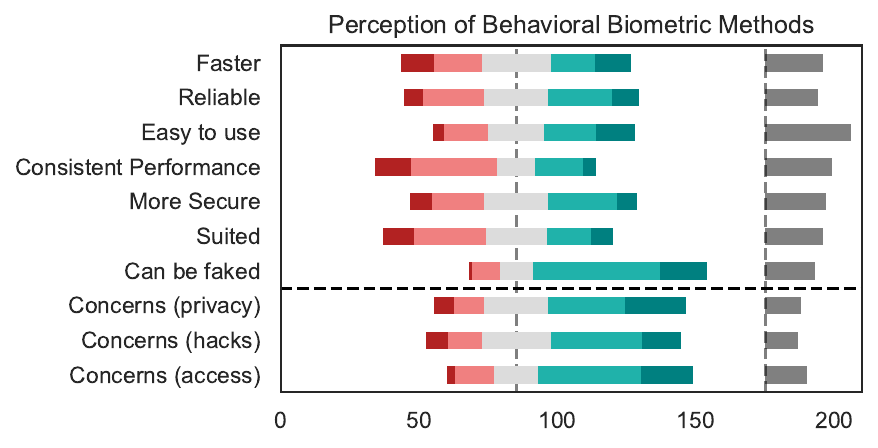}
    \caption{Participants' ratings on the Likert statements combined for both rounds of our online survey. Participants that did not give a rating are indicated in gray. See Appendix \ref{app:likerts} for the full questions. }
    \label{fig:likerts}
\end{figure*}

We also asked participants about their mobile device authentication scheme (see Table \ref{tab:smartphone} for full results). The most common combination used was fingerprint authentication ($N_{R1}=30$, $N_{R2}=23$) with a fallback PIN ($N_{R1}=28$, $N_{R2}=25$). While face recognition was not used among our participants in the first round of the survey, 12 participants (26\%) used this scheme for authentication in the second round. 

Finally, we also asked participants to rate a collection of Likert statements on the usability and security of biometrics on a scale from 1 (strongly disagree) to 5 (strongly agree). Figure \ref{fig:likerts} shows the results. We compared ratings across rounds with a Mann-Whitey U test, finding significant differences only for participants' perceptions of the consistency of biometrics. Participants in the second round disagreed with the performance of biometrics being equal for all users while they were neutral in the first session (Z = 4092.00, p = .014). 

To find potential differences in the perception of physiological and behavioral biometrics, we conducted Wilcoxon tests excluding answers where neither type of biometrics was rated. Physiological biometrics were rated significantly faster compared to Pin/Password than behavioral biometrics (Mdn$_{PB}$ = 5, Mdn$_{BB}$ = 3, Z = 5.60, p < .001). Similarly, they were rated more reliable (Mdn$_{PB}$ = 4, Mdn$_{BB}$ = 3, Z = 3.80, p < .001) and easier to use (Mdn$_{PB}$ = 5, Mdn$_{BB}$ = 3, Z = 4.96, p < .001). Participants rated physiological biometrics as significantly more secure compared to PIN/Password than behavioral biometrics (Mdn$_{PB}$ = 4, Mdn$_{BB}$ = 3, Z = 2.41, p < .014) and found them better suited to protect their personal data (Mdn$_{PB}$ = 4, Mdn$_{BB}$ = 3, Z = 4.06, p < .001). 
Concerns about privacy (Mdn$_{PB}$ = 3, Mdn$_{BB}$ = 4, Z = -3.14, p = .001), 
about being hacked (Mdn$_{PB}$ = 3, Mdn$_{BB}$ = 4, Z = -4.45, p < .001) and 
about losing access (Mdn$_{PB}$ = 3, Mdn$_{BB}$ = 4, Z = -4.45, p < .001) were all rated significantly higher for behavioral biometrics.

\subsection{Attacks and Challenges}

The majority of participants in both rounds of our survey indicated to believe that someone else could access their device when using physiological (access$_{R1}$ = 56\%,  access$_{R2}$ = 64\%) or behavioral (access$_{R1}$ = 47\%, access$_{R2}$ = 51\%) biometrics (see Table \ref{tab:quantatives}). Notably, a large group of participants indicated they did not know if this was possible -- particularly for behavioral methods (43\%  and 42\%  in the two rounds respectively). We saw very similar results regarding the effect of changes on biometric methods. A majority of participants believed that changes in their physiology or behavior would have an impact on the respective biometric methods (impact$_{R1}$ = 56\%, impact$_{R2}$ = 66\%). We did not find effects of biometric type or survey round on both of those measures when using a Fisher's exact test.

In their open-text responses, almost all participants mentioned aspects related to attacking biometrics and current challenges.
Participants saw different attack vectors for biometrics.
They mentioned software-based attacks ($N_{R1}=13$, $N_{R2}=10$), including hacking or circumventing of biometrics. They described the application of force ($N_{R1}=7$, $N_{R2}=9$), including destroyed hardware, removed body parts, or an attack during sleep (e.g. P314$_{R2}$: ``\textit{When sleeping/unconscious most of the physiological biometrics can be used without my consent}''). Participants further mentioned imitation/replay attacks ($N_{R1}=36$, $N_{R2}=36$), including deepfakes and mimicry attacks (e.g. P328$_{R2}$: ``\textit{trying to build an imprint of a fingerprint or (more elaborate): building a facemask}''), and gave some other attack vectors ($N_{R1}=11$, $N_{R2}=5$), including social engineering or outsourcing attacks (e.g. P21$_{R1}$: ``\textit{I would pay someone to do it for me}'').

Many participants also mentioned non-malicious reasons why biometrics could fail.
They gave generic reasons ($N_{R1}=18$, $N_{R2}=23$), such as changed hardware or a longer time span between authentication attempts and physical reasons ($N_{R1}=20$, $N_{R2}=24$), such as cosmetics, haircuts, or injuries. For example P297$_{R2}$ named ``\textit{diseases of skin [and] injuries}'' as potential reasons for fingerprint not working, adding they had personal experience with non-detection from ``\textit{very dry skin with deep cuts from outdoor work}''. Finally, participants mentioned behavioral reasons ($N_{R1}=2$, $N_{R2}=7$), such as an impact resulting from mood or a purposefully changed behavior (e.g. P16$_{R1}$: ``\textit{I type much faster when I argue with my girlfriend. This might bias the system}''). 

Finally, many participants expressed perceived weaknesses of biometrics that exist from their perspectives ($N_{R1}=29$, $N_{R2}=20$), such as the possibility of faking them or the collection of their identity by governments or companies.
A few participants stated that they do not believe that biometrics can be attacked or it would not have an impact ($N_{R1}=9$, $N_{R2}=4$). For example, as an answer to whether physiological changes could impact biometrics, P98$_{R1}$ answered: ``\textit{Not for fingerprints}''.

\subsection{Consequences of an Attack}
Most participants explained how they would deal with a successful attack on their biometric authentication and voiced potential damage resulting from such an attack.
In case of a successful attack, participants highlighted different actions they would take.
Most participants stated that they would fall back to another authentication method ($N_{R1}=50$, $N_{R2}=43$), while several participants stated they would try to control the damage ($N_{R1}=24$, $N_{R2}=21$), by, for example, informing contacts, recovering data or resetting their device. A different strategy was seeking support ($N_{R1}=24$, $N_{R2}=10$), by contacting the police or their provider (e.g. P67$_{R1}$: ``\textit{I would research the right institution to call in this case or go to the police}'').
Concerning the potential damage resulting from a successful attack, participants mention that the attacker could misuse (e.g. P124$_{R1}$: ``\textit{I would be unable to prove that this person is not me}'') or simply access their data ($N_{R1}=17$, $N_{R2}=18$) or that they would lose their data ($N_{R1}=17$, $N_{R2}=11$). A few participants stated that an attack would have no impact or they did not intend to react to it ($N_{R1}=9$, $N_{R2}=4$).

\subsection{Future Suggestions \& Improvements}
Finally, participants listed different suggestions that could improve biometric systems in the future.
Many participants mention novel forms of biometrics ($N_{R1}=37$, $N_{R2}=34$), including body and facial movements. P343$_{R2}$ suggested using ``\textit{genetics through noninvasive tissue sampling (high level of security should sequencing in real time become possible)}''.
Moreover, participants stated new applications for which biometrics can be used in the future ($N_{R1}=7$, $N_{R2}=5$), including payments (e.g. P19$_{R1}$:``\textit{Maybe payment with DNA identification}''), forensics, public contexts, and sports.
Finally, many participants gave general improvements for biometrics ($N_{R1}=27$, $N_{R2}=30$), including tweaking thresholds, updating models with more data, increasing the precision of sensors, and use hard to replicate methods. Finally, another often mentioned improvement was ``\textit{using more than one type of [physiological or] behavioral biometric system}'' (P304$_{R2}$).

\section{Discussion}

\subsection{Do they understand what they are using?}
Many participants indicated to be unfamiliar with biometric methods or were unable to define or explain the concept. 
Nonetheless, a larger number of them were able to at least give examples, name features used, or correctly associate them with authentication. Thus, asking open questions in addition to the self-assessment allowed us to see that many participants seemed to have at least some knowledge about biometrics and answers often included important aspects of the definition we used. A related study found many participants to be aware of face recognition~\cite{ada2019beyond}, but only a few had deeper knowledge. We did not focus on face recognition but observed a similar effect for biometrics in general. 

However, some participants had only an abstract association (e.g. from having heard the term in the context of their passport).
We also received many answers that were not related to biometric authentication but referred to, for example, body functions, health, or influences on behavior.
This can mean one of two things: either participants were indeed not familiar with biometrics or they just had a different association with the terms used. 

While it is unclear to which degree participants need a detailed understanding of biometric methods in order to confidently and securely employ them in their daily lives we argue that a better understanding of the risks and potential shortcomings could ultimately foster safe use and trust. 
%
Understanding what features are collected and how they are used is essential for informed consent to use those approaches. Misconceptions and missing knowledge on the other hand can lead to users needlessly abandoning biometrics, not adopting them in the first place, or putting their data at risk (e.g. due to ignoring leaked physiological traits like fingerprints or not retraining a behavioral model). 

Regarding the term `biometrics' itself, one participant explicitly preferred the term anatomical over physiological biometrics. More generally, we believe that many participants may have made a connection from biometrics to \textit{bio}logy and \textit{metrics} and thus assumed a connection to e.g. body functions rather than security. Potentially, renaming physiological biometrics to e.g. appearance-based identification could aid in clarifying their function. 

\subsection{How do they cope with problems?}

While many participants had very sophisticated ideas about how biometric systems could be attacked, surprisingly few participants mentioned attacking the fallback method. However, every modern biometric system uses a fallback like a PIN to enable access in case the biometric factor fails. Many participants stated they believe biometrics to be more secure than using those traditional methods but practically, this currently cannot be the case. This is something that biometric interfaces should clearly communicate to users, for example during enrollment. 

At the same time about half of our participants thought that someone else could access their device if protected by a biometric system; showing a contradicting tendency. 
Many indicated they would switch from biometrics to other authentication methods in case they experienced issues with changes (e.g. in their appearance) or were to be attacked. Others had no idea how to cope with problems and suggested outsourcing their solution, e.g. to family members. 

Overall, this means we observed two opposite tendencies, with some participants believing that biometrics could not be attacked and being very confident in their security while others mistrusted the technology and had no clear plan for how to handle issues should they come up. 
As always, the truth lies somewhere in between. Education may help users to gain a more reflected impression of biometrics, avoid a false sense of security, and prepare them for potential issues. 

\subsection{Do they know behavioral biometrics?}

In contrast to the majority of related work, we actively distinguished physiological and behavioral biometrics to understand how perception and understanding differed between the two. While many participants could name physiological biometrics, the only behavioral method mentioned by more than 20\% of the participants in both rounds was voice recognition. Similarly, more participants did not know if changes would affect behavioral systems and if an attacker could gain access to such a system. 
Participants had significantly more concerns about behavioral biometrics and rated them slower, less reliable, and harder to use than their physiological counterparts. 

All those aspects imply that knowledge of behavioral methods is less prevalent among our participants, leading to increased uncertainty. This is in line with previous work~\cite{furnell2007public,buckley2019language} and intuitively makes sense, as behavioral methods are normally designed to be transparent and facilitate authentication without explicit action or even knowledge of the user. Yet, this is also a risk, as behavioral traits can be used for (unwanted) profiling and tracking purposes (e.g. recognizing users on a website by their mouse movements~\cite{pusara2004user} or in public places by their gait~\cite{vera2013gait}) even though behavioral biometrics are not (yet) widely used.

\subsection{Did something change?}
We designed our study to find potential differences in the use and perception of biometrics over time. To achieve this we openly coded the first round and used the resulting codebook on the answers from the second round to be able to find newly emerging themes and codes. Given the increasing adoption of biometrics in users' daily lives over the recent years we would have expected to see a change in users' perception and knowledge between our rounds. However, this was not the case: no codes emerged or disappeared between rounds (Appendix \ref{app:codebook}). 

Our quantitative results show that the usage of biometrics as the primary authentication mechanism on smartphones increased from 53\% in the first round of our survey to 75\% in the second round, i.e. three out of four of our participants indicated using either fingerprint or face recognition. Participants indicated significantly less agreement with the performance of biometrics being equal for all users, which may hint at an increased awareness of biases in those models. However, this is speculation so far and would need further confirmation in a future study. 
Apart from that -- and again contrary to our expectation -- the increased adoption did not seem to have led to significant differences in perception. For instance, we did not observe changes in perceived ease of use, reliability, or general concerns. Also, participants still expressed (similar) concerns with regard to the consistency and security of biometric mechanisms.
We further believe in the approach of comparing data over a longer time frame, though it may be worthwhile to also explore other methods (e.g. using more closed questions or conducting interviews based on our findings) to better understand such effects.

\subsection{Practical Implications}

In our results, we uncovered several gaps in participants' knowledge that may lead to unsafe use of biometric authentication. Here, we suggest two directions for designers to address this in future interfaces for biometric systems. 

The first starting point would be to foster user \textit{literacy} and awareness with regard to biometrics. This can, e.g., be done when setting up the authentication by clearly communicating that biometrics can offer additional convenience but do not provide additional security as long as a weaker fallback exists. If changes in the users' physiology or appearance occur over time (e.g. changed hairstyle or walking patterns), a system can also inform users about such changes and their potential consequences for authentication. This can help users to build a better mental model of how their authentication works as well as prompt them to trigger potential steps to solve the issue, like retraining.
Literacy is particularly important for behavioral biometrics as they are (as of now) often not used by end-users but \textit{on} them, e.g. in the form of tracking or profiling. Here it is important to understand how those mechanisms work and how collected data can expose users' identity (e.g. online or in public). 

The second point that can be addressed is giving \textit{control} to the user. A prerequisite for informed use of such control options is the literacy mentioned above. 
This can entail fine-grained control over the data collected (e.g. during setup), when this data is collected (e.g. no gait tracking when at home), and with whom it is shared (e.g. a legitimate app the user wants to use or third parties). Another direction would be supporting users in protecting their identity in contexts where they might be unwillingly identified by biometric methods. Examples could be prompting users to walk differently in a known location with gait recognition or by obfuscating typing behavior by adding random delays. 

\section{Conclusion}

In this paper, we presented our investigation of user perception of physiological and behavioral biometrics. We conducted a survey in two rounds and derived themes from participants' responses. Our results show, that most participants actively use biometric methods in their daily life and prefer them over traditional methods. At the same time, most participants lacked in-depth knowledge about how biometrics work and showed uncertainties with regard to handling potential problems caused by an attack or changes in their physiology or behavior. Here we see potential for improvement through further research on users' perception and understanding of biometrics and the design of biometric interfaces that foster their literacy of biometric methods and give them more control over them.

Between the two rounds of our survey, we saw a strong increase in the adoption of biometric methods and it is very plausible that this trend will continue with improving algorithms and sensors. Thus, the time to improve user interaction with biometric systems is now to enable future safe and informed use of this technology.

\begin{acks}
This project is funded by the Deutsche Forschungsgemeinschaft
(DFG, German Research Foundation) – 425869382 and is part of
Priority Program SPP2199 Scalable Interaction Paradigms for Pervasive Computing Environments.
\end{acks}

\bibliographystyle{ACM-Reference-Format}
\bibliography{bib}


\begin{thebibliography}{64}


\ifx \showCODEN    \undefined \def \showCODEN     #1{\unskip}     \fi
\ifx \showDOI      \undefined \def \showDOI       #1{#1}\fi
\ifx \showISBNx    \undefined \def \showISBNx     #1{\unskip}     \fi
\ifx \showISBNxiii \undefined \def \showISBNxiii  #1{\unskip}     \fi
\ifx \showISSN     \undefined \def \showISSN      #1{\unskip}     \fi
\ifx \showLCCN     \undefined \def \showLCCN      #1{\unskip}     \fi
\ifx \shownote     \undefined \def \shownote      #1{#1}          \fi
\ifx \showarticletitle \undefined \def \showarticletitle #1{#1}   \fi
\ifx \showURL      \undefined \def \showURL       {\relax}        \fi
\providecommand\bibfield[2]{#2}
\providecommand\bibinfo[2]{#2}
\providecommand\natexlab[1]{#1}
\providecommand\showeprint[2][]{arXiv:#2}

\bibitem[Adams and Sasse(1999)]%
        {adams1999}
\bibfield{author}{\bibinfo{person}{Anne Adams} {and}
  \bibinfo{person}{Martina~Angela Sasse}.} \bibinfo{year}{1999}\natexlab{}.
\newblock \showarticletitle{Users Are Not the Enemy}.
\newblock \bibinfo{journal}{\emph{Commun. ACM}} \bibinfo{volume}{42},
  \bibinfo{number}{12} (\bibinfo{date}{dec} \bibinfo{year}{1999}),
  \bibinfo{pages}{40–46}.
\newblock
\showISSN{0001-0782}
\urldef\tempurl%
\url{https://doi.org/10.1145/322796.322806}
\showDOI{\tempurl}


\bibitem[Alhussain et~al\mbox{.}(2013)]%
        {alhussain2013users}
\bibfield{author}{\bibinfo{person}{Thamer Alhussain}, \bibinfo{person}{Rayed
  AlGhamdi}, \bibinfo{person}{Salem Alkhalaf}, {and} \bibinfo{person}{Osama
  Alfarraj}.} \bibinfo{year}{2013}\natexlab{}.
\newblock \showarticletitle{{Users' Perceptions of Mobile Phone Security: A
  Survey Study in the Kingdom of Saudi Arabia}}.
\newblock \bibinfo{journal}{\emph{international journal of computer theory and
  engineering}} \bibinfo{volume}{5}, \bibinfo{number}{5}
  (\bibinfo{year}{2013}), \bibinfo{pages}{793}.
\newblock


\bibitem[Asgharpour et~al\mbox{.}(2007)]%
        {asgharpour2007}
\bibfield{author}{\bibinfo{person}{Farzaneh Asgharpour}, \bibinfo{person}{Debin
  Liu}, {and} \bibinfo{person}{L.~Jean Camp}.} \bibinfo{year}{2007}\natexlab{}.
\newblock \showarticletitle{Mental Models of Security Risks}. In
  \bibinfo{booktitle}{\emph{Financial Cryptography and Data Security}},
  \bibfield{editor}{\bibinfo{person}{Sven Dietrich} {and}
  \bibinfo{person}{Rachna Dhamija}} (Eds.). \bibinfo{publisher}{Springer Berlin
  Heidelberg}, \bibinfo{address}{Berlin, Heidelberg},
  \bibinfo{pages}{367--377}.
\newblock
\showISBNx{978-3-540-77366-5}


\bibitem[Bhagavatula et~al\mbox{.}(2015)]%
        {bhagavatula2015biometric}
\bibfield{author}{\bibinfo{person}{Rasekhar Bhagavatula},
  \bibinfo{person}{Blase Ur}, \bibinfo{person}{Kevin Iacovino},
  \bibinfo{person}{Su~Mon Kywe}, \bibinfo{person}{Lorrie~Faith Cranor}, {and}
  \bibinfo{person}{Marios Savvides}.} \bibinfo{year}{2015}\natexlab{}.
\newblock \showarticletitle{{Biometric authentication on iphone and android:
  Usability, perceptions, and influences on adoption}}.
\newblock  (\bibinfo{year}{2015}).
\newblock


\bibitem[Bharathi and Shamily(2020)]%
        {bharathi2020review}
\bibfield{author}{\bibinfo{person}{B Bharathi} {and} \bibinfo{person}{P~Bindhu
  Shamily}.} \bibinfo{year}{2020}\natexlab{}.
\newblock \showarticletitle{A review on iris recognition system for person
  identification}.
\newblock \bibinfo{journal}{\emph{International Journal of Computational
  Biology and Drug Design}} \bibinfo{volume}{13}, \bibinfo{number}{3}
  (\bibinfo{year}{2020}), \bibinfo{pages}{316--331}.
\newblock


\bibitem[Braun and Clarke(2021)]%
        {braun2021can}
\bibfield{author}{\bibinfo{person}{Virginia Braun} {and}
  \bibinfo{person}{Victoria Clarke}.} \bibinfo{year}{2021}\natexlab{}.
\newblock \showarticletitle{{Can I use TA? Should I use TA? Should I not use
  TA? Comparing reflexive thematic analysis and other pattern-based qualitative
  analytic approaches}}.
\newblock \bibinfo{journal}{\emph{Counselling and Psychotherapy Research}}
  \bibinfo{volume}{21}, \bibinfo{number}{1} (\bibinfo{year}{2021}),
  \bibinfo{pages}{37--47}.
\newblock


\bibitem[Buckley and Nurse(2019)]%
        {buckley2019language}
\bibfield{author}{\bibinfo{person}{Oliver Buckley} {and}
  \bibinfo{person}{Jason~RC Nurse}.} \bibinfo{year}{2019}\natexlab{}.
\newblock \showarticletitle{The language of biometrics: Analysing public
  perceptions}.
\newblock \bibinfo{journal}{\emph{Journal of Information Security and
  Applications}}  \bibinfo{volume}{47} (\bibinfo{year}{2019}),
  \bibinfo{pages}{112--119}.
\newblock


\bibitem[Bulling et~al\mbox{.}(2012)]%
        {bulling2012increasing}
\bibfield{author}{\bibinfo{person}{Andreas Bulling}, \bibinfo{person}{Florian
  Alt}, {and} \bibinfo{person}{Albrecht Schmidt}.}
  \bibinfo{year}{2012}\natexlab{}.
\newblock \showarticletitle{Increasing the security of gaze-based cued-recall
  graphical passwords using saliency masks}. In
  \bibinfo{booktitle}{\emph{Proceedings of the SIGCHI Conference on Human
  Factors in Computing Systems}}. \bibinfo{pages}{3011--3020}.
\newblock


\bibitem[Buschek et~al\mbox{.}(2015)]%
        {buschek2015}
\bibfield{author}{\bibinfo{person}{Daniel Buschek}, \bibinfo{person}{Alexander
  De~Luca}, {and} \bibinfo{person}{Florian Alt}.}
  \bibinfo{year}{2015}\natexlab{}.
\newblock \showarticletitle{Improving Accuracy, Applicability and Usability of
  Keystroke Biometrics on Mobile Touchscreen Devices}. In
  \bibinfo{booktitle}{\emph{Proceedings of the 33rd Annual ACM Conference on
  Human Factors in Computing Systems}} (Seoul, Republic of Korea)
  \emph{(\bibinfo{series}{CHI '15})}. \bibinfo{publisher}{Association for
  Computing Machinery}, \bibinfo{address}{New York, NY, USA},
  \bibinfo{pages}{1393–1402}.
\newblock
\showISBNx{9781450331456}
\urldef\tempurl%
\url{https://doi.org/10.1145/2702123.2702252}
\showDOI{\tempurl}


\bibitem[Byun and Byun(2013)]%
        {byun2013exploring}
\bibfield{author}{\bibinfo{person}{Sookeun Byun} {and}
  \bibinfo{person}{Sang-Eun Byun}.} \bibinfo{year}{2013}\natexlab{}.
\newblock \showarticletitle{Exploring perceptions toward biometric technology
  in service encounters: a comparison of current users and potential adopters}.
\newblock \bibinfo{journal}{\emph{Behaviour \& Information Technology}}
  \bibinfo{volume}{32}, \bibinfo{number}{3} (\bibinfo{year}{2013}),
  \bibinfo{pages}{217--230}.
\newblock


\bibitem[Dantcheva et~al\mbox{.}(2015)]%
        {dantcheva2015else}
\bibfield{author}{\bibinfo{person}{Antitza Dantcheva}, \bibinfo{person}{Petros
  Elia}, {and} \bibinfo{person}{Arun Ross}.} \bibinfo{year}{2015}\natexlab{}.
\newblock \showarticletitle{What else does your biometric data reveal? A survey
  on soft biometrics}.
\newblock \bibinfo{journal}{\emph{IEEE Transactions on Information Forensics
  and Security}} \bibinfo{volume}{11}, \bibinfo{number}{3}
  (\bibinfo{year}{2015}), \bibinfo{pages}{441--467}.
\newblock


\bibitem[Dargan and Kumar(2020)]%
        {dargan2020comprehensive}
\bibfield{author}{\bibinfo{person}{Shaveta Dargan} {and}
  \bibinfo{person}{Munish Kumar}.} \bibinfo{year}{2020}\natexlab{}.
\newblock \showarticletitle{A comprehensive survey on the biometric recognition
  systems based on physiological and behavioral modalities}.
\newblock \bibinfo{journal}{\emph{Expert Systems with Applications}}
  \bibinfo{volume}{143} (\bibinfo{year}{2020}), \bibinfo{pages}{113114}.
\newblock


\bibitem[De~Marsico et~al\mbox{.}(2016)]%
        {de2016iris}
\bibfield{author}{\bibinfo{person}{Maria De~Marsico}, \bibinfo{person}{Alfredo
  Petrosino}, {and} \bibinfo{person}{Stefano Ricciardi}.}
  \bibinfo{year}{2016}\natexlab{}.
\newblock \showarticletitle{Iris recognition through machine learning
  techniques: A survey}.
\newblock \bibinfo{journal}{\emph{Pattern Recognition Letters}}
  \bibinfo{volume}{82} (\bibinfo{year}{2016}), \bibinfo{pages}{106--115}.
\newblock


\bibitem[Drozdowski et~al\mbox{.}(2020)]%
        {drozdowski2020demographic}
\bibfield{author}{\bibinfo{person}{Pawel Drozdowski},
  \bibinfo{person}{Christian Rathgeb}, \bibinfo{person}{Antitza Dantcheva},
  \bibinfo{person}{Naser Damer}, {and} \bibinfo{person}{Christoph Busch}.}
  \bibinfo{year}{2020}\natexlab{}.
\newblock \showarticletitle{Demographic bias in biometrics: A survey on an
  emerging challenge}.
\newblock \bibinfo{journal}{\emph{IEEE Transactions on Technology and Society}}
  \bibinfo{volume}{1}, \bibinfo{number}{2} (\bibinfo{year}{2020}),
  \bibinfo{pages}{89--103}.
\newblock


\bibitem[Ellavarason et~al\mbox{.}(2020)]%
        {ellavarason2020touch}
\bibfield{author}{\bibinfo{person}{Elakkiya Ellavarason},
  \bibinfo{person}{Richard Guest}, \bibinfo{person}{Farzin Deravi},
  \bibinfo{person}{Raul Sanchez-Riello}, {and} \bibinfo{person}{Barbara
  Corsetti}.} \bibinfo{year}{2020}\natexlab{}.
\newblock \showarticletitle{Touch-dynamics based behavioural biometrics on
  mobile devices--a review from a usability and performance perspective}.
\newblock \bibinfo{journal}{\emph{ACM Computing Surveys (CSUR)}}
  \bibinfo{volume}{53}, \bibinfo{number}{6} (\bibinfo{year}{2020}),
  \bibinfo{pages}{1--36}.
\newblock


\bibitem[Elliott et~al\mbox{.}(2007)]%
        {elliott2007perception}
\bibfield{author}{\bibinfo{person}{Stephen~J Elliott}, \bibinfo{person}{Sarah~A
  Massie}, {and} \bibinfo{person}{Mathias~J Sutton}.}
  \bibinfo{year}{2007}\natexlab{}.
\newblock \showarticletitle{{The perception of biometric technology: A
  survey}}. In \bibinfo{booktitle}{\emph{Automatic Identification Advanced
  Technologies, 2007 IEEE Workshop on}}. IEEE, \bibinfo{pages}{259--264}.
\newblock


\bibitem[Franks and Smith(2020)]%
        {franks2020identity}
\bibfield{author}{\bibinfo{person}{Christie Franks} {and}
  \bibinfo{person}{Russell~G Smith}.} \bibinfo{year}{2020}\natexlab{}.
\newblock \showarticletitle{{Identity crime and misuse in Australia: Results of
  the 2019 online survey}}.
\newblock  (\bibinfo{year}{2020}).
\newblock


\bibitem[Franks and Smith(2021)]%
        {franks2021changing}
\bibfield{author}{\bibinfo{person}{Christie Franks} {and}
  \bibinfo{person}{Russell~G Smith}.} \bibinfo{year}{2021}\natexlab{}.
\newblock \bibinfo{booktitle}{\emph{Changing perceptions of biometric
  technologies}}.
\newblock \bibinfo{publisher}{Australian Institute of Criminology}.
\newblock


\bibitem[Furnell and Evangelatos(2007)]%
        {furnell2007public}
\bibfield{author}{\bibinfo{person}{Steven Furnell} {and}
  \bibinfo{person}{Konstantinos Evangelatos}.} \bibinfo{year}{2007}\natexlab{}.
\newblock \showarticletitle{{Public awareness and perceptions of biometrics}}.
\newblock \bibinfo{journal}{\emph{Computer Fraud \& Security}}
  \bibinfo{volume}{2007}, \bibinfo{number}{1} (\bibinfo{year}{2007}),
  \bibinfo{pages}{8--13}.
\newblock


\bibitem[Gafurov(2007)]%
        {gafurov2007survey}
\bibfield{author}{\bibinfo{person}{Davrondzhon Gafurov}.}
  \bibinfo{year}{2007}\natexlab{}.
\newblock \showarticletitle{A survey of biometric gait recognition: Approaches,
  security and challenges}. In \bibinfo{booktitle}{\emph{Annual Norwegian
  computer science conference}}. Annual Norwegian Computer Science Conference
  Norway, \bibinfo{pages}{19--21}.
\newblock


\bibitem[Galar et~al\mbox{.}(2015)]%
        {galar2015survey}
\bibfield{author}{\bibinfo{person}{Mikel Galar}, \bibinfo{person}{Joaqu{\'\i}n
  Derrac}, \bibinfo{person}{Daniel Peralta}, \bibinfo{person}{Isaac Triguero},
  \bibinfo{person}{Daniel Paternain}, \bibinfo{person}{Carlos Lopez-Molina},
  \bibinfo{person}{Salvador Garc{\'\i}a}, \bibinfo{person}{Jos{\'e}~M
  Ben{\'\i}tez}, \bibinfo{person}{Miguel Pagola}, \bibinfo{person}{Edurne
  Barrenechea}, {et~al\mbox{.}}} \bibinfo{year}{2015}\natexlab{}.
\newblock \showarticletitle{A survey of fingerprint classification Part I:
  Taxonomies on feature extraction methods and learning models}.
\newblock \bibinfo{journal}{\emph{Knowledge-based systems}}
  \bibinfo{volume}{81} (\bibinfo{year}{2015}), \bibinfo{pages}{76--97}.
\newblock


\bibitem[Harboe and Huang(2015)]%
        {harboe2015realworld}
\bibfield{author}{\bibinfo{person}{Gunnar Harboe} {and}
  \bibinfo{person}{Elaine~M. Huang}.} \bibinfo{year}{2015}\natexlab{}.
\newblock \showarticletitle{{Real-World Affinity Diagramming Practices:
  Bridging the Paper-Digital Gap}}. In \bibinfo{booktitle}{\emph{Proc. 33rd
  Annual ACM Conf. Human Factors in Computing Systems}}.
  \bibinfo{publisher}{ACM}, \bibinfo{address}{New York, NY, USA},
  \bibinfo{pages}{95--104}.
\newblock
\showISBNx{978-1-4503-3145-6}
\urldef\tempurl%
\url{https://doi.org/10.1145/2702123.2702561}
\showDOI{\tempurl}


\bibitem[Institute(2019)]%
        {ada2019beyond}
\bibfield{author}{\bibinfo{person}{Ada~Lovelace Institute}.}
  \bibinfo{year}{2019}\natexlab{}.
\newblock \showarticletitle{{Beyond Face Value: Public Attitudes to Facial
  Recognition Technology.}}
\newblock  (\bibinfo{year}{2019}).
\newblock


\bibitem[ISO/IEC 2382-37(2012)]%
        {iso2382-37}
ISO/IEC 2382-37 \bibinfo{year}{2012}\natexlab{}.
\newblock \bibinfo{booktitle}{\emph{{Biometric Vocabulary}}}.
\newblock \bibinfo{type}{{T}echnical {R}eport}.
  \bibinfo{institution}{International Organization for Standardization(ISO)}.
\newblock


\bibitem[Jain et~al\mbox{.}(1996)]%
        {jain1996introduction}
\bibfield{author}{\bibinfo{person}{Anil Jain}, \bibinfo{person}{Ruud Bolle},
  {and} \bibinfo{person}{Sharath Pankanti}.} \bibinfo{year}{1996}\natexlab{}.
\newblock \bibinfo{booktitle}{\emph{Introduction to biometrics}}.
\newblock \bibinfo{publisher}{Springer}.
\newblock


\bibitem[Jain et~al\mbox{.}(1999)]%
        {jain1999multichannel}
\bibfield{author}{\bibinfo{person}{A.K. Jain}, \bibinfo{person}{S. Prabhakar},
  {and} \bibinfo{person}{Lin Hong}.} \bibinfo{year}{1999}\natexlab{}.
\newblock \showarticletitle{A multichannel approach to fingerprint
  classification}.
\newblock \bibinfo{journal}{\emph{IEEE Transactions on Pattern Analysis and
  Machine Intelligence}} \bibinfo{volume}{21}, \bibinfo{number}{4}
  (\bibinfo{year}{1999}), \bibinfo{pages}{348--359}.
\newblock
\urldef\tempurl%
\url{https://doi.org/10.1109/34.761265}
\showDOI{\tempurl}


\bibitem[Jain and Kumar(2010)]%
        {jain2010biometrics}
\bibfield{author}{\bibinfo{person}{Anil~K Jain} {and} \bibinfo{person}{Ajay
  Kumar}.} \bibinfo{year}{2010}\natexlab{}.
\newblock \showarticletitle{Biometrics of next generation: An overview}.
\newblock \bibinfo{journal}{\emph{Second generation biometrics}}
  \bibinfo{volume}{12}, \bibinfo{number}{1} (\bibinfo{year}{2010}),
  \bibinfo{pages}{2--3}.
\newblock


\bibitem[Jain et~al\mbox{.}(2004)]%
        {jain2004introduction}
\bibfield{author}{\bibinfo{person}{Anil~K Jain}, \bibinfo{person}{Arun Ross},
  {and} \bibinfo{person}{Salil Prabhakar}.} \bibinfo{year}{2004}\natexlab{}.
\newblock \showarticletitle{An introduction to biometric recognition}.
\newblock \bibinfo{journal}{\emph{IEEE Transactions on circuits and systems for
  video technology}} \bibinfo{volume}{14}, \bibinfo{number}{1}
  (\bibinfo{year}{2004}), \bibinfo{pages}{4--20}.
\newblock


\bibitem[Karatzouni et~al\mbox{.}(2007)]%
        {karatzouni2007perceptions}
\bibfield{author}{\bibinfo{person}{Sevasti Karatzouni},
  \bibinfo{person}{Steven~M Furnell}, \bibinfo{person}{Nathan~L Clarke}, {and}
  \bibinfo{person}{Reinhardt~A Botha}.} \bibinfo{year}{2007}\natexlab{}.
\newblock \showarticletitle{Perceptions of user authentication on mobile
  devices}. In \bibinfo{booktitle}{\emph{Proceedings of the ISOneWorld
  Conference}}. Citeseer, \bibinfo{pages}{11--13}.
\newblock


\bibitem[Karnan et~al\mbox{.}(2011)]%
        {karnan2011biometric}
\bibfield{author}{\bibinfo{person}{Marcus Karnan},
  \bibinfo{person}{Muthuramalingam Akila}, {and} \bibinfo{person}{Nishara
  Krishnaraj}.} \bibinfo{year}{2011}\natexlab{}.
\newblock \showarticletitle{Biometric personal authentication using keystroke
  dynamics: A review}.
\newblock \bibinfo{journal}{\emph{Applied soft computing}}
  \bibinfo{volume}{11}, \bibinfo{number}{2} (\bibinfo{year}{2011}),
  \bibinfo{pages}{1565--1573}.
\newblock


\bibitem[Kong et~al\mbox{.}(2009)]%
        {kong2009survey}
\bibfield{author}{\bibinfo{person}{Adams Kong}, \bibinfo{person}{David Zhang},
  {and} \bibinfo{person}{Mohamed Kamel}.} \bibinfo{year}{2009}\natexlab{}.
\newblock \showarticletitle{A survey of palmprint recognition}.
\newblock \bibinfo{journal}{\emph{pattern recognition}} \bibinfo{volume}{42},
  \bibinfo{number}{7} (\bibinfo{year}{2009}), \bibinfo{pages}{1408--1418}.
\newblock


\bibitem[Liebers and Schneegass(2020)]%
        {liebers2020introducing}
\bibfield{author}{\bibinfo{person}{Jonathan Liebers} {and}
  \bibinfo{person}{Stefan Schneegass}.} \bibinfo{year}{2020}\natexlab{}.
\newblock \showarticletitle{Introducing functional biometrics: Using
  body-reflections as a novel class of biometric authentication systems}. In
  \bibinfo{booktitle}{\emph{Extended Abstracts of the 2020 CHI Conference on
  Human Factors in Computing Systems}}. \bibinfo{pages}{1--7}.
\newblock


\bibitem[Maltoni et~al\mbox{.}(2009)]%
        {maltoni2009handbook}
\bibfield{author}{\bibinfo{person}{Davide Maltoni}, \bibinfo{person}{Dario
  Maio}, \bibinfo{person}{Anil~K Jain}, \bibinfo{person}{Salil Prabhakar},
  {et~al\mbox{.}}} \bibinfo{year}{2009}\natexlab{}.
\newblock \bibinfo{booktitle}{\emph{Handbook of fingerprint recognition}}.
  Vol.~\bibinfo{volume}{2}.
\newblock \bibinfo{publisher}{Springer}.
\newblock


\bibitem[Marasco and Ross(2014)]%
        {marasco2014survey}
\bibfield{author}{\bibinfo{person}{Emanuela Marasco} {and}
  \bibinfo{person}{Arun Ross}.} \bibinfo{year}{2014}\natexlab{}.
\newblock \showarticletitle{A survey on antispoofing schemes for fingerprint
  recognition systems}.
\newblock \bibinfo{journal}{\emph{ACM Computing Surveys (CSUR)}}
  \bibinfo{volume}{47}, \bibinfo{number}{2} (\bibinfo{year}{2014}),
  \bibinfo{pages}{1--36}.
\newblock


\bibitem[McDonald et~al\mbox{.}(2019)]%
        {McDonald2019reliability}
\bibfield{author}{\bibinfo{person}{Nora McDonald}, \bibinfo{person}{Sarita
  Schoenebeck}, {and} \bibinfo{person}{Andrea Forte}.}
  \bibinfo{year}{2019}\natexlab{}.
\newblock \showarticletitle{{Reliability and Inter-Rater Reliability in
  Qualitative Research: Norms and Guidelines for CSCW and HCI Practice}}.
\newblock \bibinfo{journal}{\emph{Proc. ACM Hum.-Comput. Interact.}}
  \bibinfo{volume}{3}, \bibinfo{number}{CSCW}, Article \bibinfo{articleno}{72}
  (\bibinfo{date}{nov} \bibinfo{year}{2019}), \bibinfo{numpages}{23}~pages.
\newblock
\urldef\tempurl%
\url{https://doi.org/10.1145/3359174}
\showDOI{\tempurl}


\bibitem[Phillips et~al\mbox{.}(2017)]%
        {phillips2017lessons}
\bibfield{author}{\bibinfo{person}{P~Jonathon Phillips},
  \bibinfo{person}{Patrick~J Flynn}, {and} \bibinfo{person}{Kevin~W Bowyer}.}
  \bibinfo{year}{2017}\natexlab{}.
\newblock \showarticletitle{Lessons from collecting a million biometric
  samples}.
\newblock \bibinfo{journal}{\emph{Image and Vision Computing}}
  \bibinfo{volume}{58} (\bibinfo{year}{2017}), \bibinfo{pages}{96--107}.
\newblock


\bibitem[Pusara and Brodley(2004)]%
        {pusara2004user}
\bibfield{author}{\bibinfo{person}{Maja Pusara} {and} \bibinfo{person}{Carla~E.
  Brodley}.} \bibinfo{year}{2004}\natexlab{}.
\newblock \showarticletitle{User Re-Authentication via Mouse Movements}. In
  \bibinfo{booktitle}{\emph{Proceedings of the 2004 ACM Workshop on
  Visualization and Data Mining for Computer Security}} (Washington DC, USA)
  \emph{(\bibinfo{series}{VizSEC/DMSEC '04})}. \bibinfo{publisher}{Association
  for Computing Machinery}, \bibinfo{address}{New York, NY, USA},
  \bibinfo{pages}{1–8}.
\newblock
\showISBNx{1581139748}
\urldef\tempurl%
\url{https://doi.org/10.1145/1029208.1029210}
\showDOI{\tempurl}


\bibitem[Rasnayaka and Sim(2018)]%
        {rasnayaka2018wants}
\bibfield{author}{\bibinfo{person}{Sanka Rasnayaka} {and}
  \bibinfo{person}{Terence Sim}.} \bibinfo{year}{2018}\natexlab{}.
\newblock \showarticletitle{Who wants continuous authentication on mobile
  devices?}. In \bibinfo{booktitle}{\emph{2018 IEEE 9th International
  Conference on Biometrics Theory, Applications and Systems (BTAS)}}. IEEE,
  \bibinfo{pages}{1--9}.
\newblock


\bibitem[Raul et~al\mbox{.}(2020)]%
        {raul2020comprehensive}
\bibfield{author}{\bibinfo{person}{Nataasha Raul}, \bibinfo{person}{Radha
  Shankarmani}, {and} \bibinfo{person}{Padmaja Joshi}.}
  \bibinfo{year}{2020}\natexlab{}.
\newblock \showarticletitle{A comprehensive review of keystroke dynamics-based
  authentication mechanism}. In \bibinfo{booktitle}{\emph{International
  Conference on Innovative Computing and Communications: Proceedings of ICICC
  2019, Volume 2}}. Springer, \bibinfo{pages}{149--162}.
\newblock


\bibitem[Saad et~al\mbox{.}(2022)]%
        {saad2022mask}
\bibfield{author}{\bibinfo{person}{Alia Saad}, \bibinfo{person}{Uwe
  Gruenefeld}, \bibinfo{person}{Lukas Mecke}, \bibinfo{person}{Marion Koelle},
  \bibinfo{person}{Florian Alt}, {and} \bibinfo{person}{Stefan Schneegass}.}
  \bibinfo{year}{2022}\natexlab{}.
\newblock \showarticletitle{{Mask removal isn’t always convenient in
  public!--The Impact of the Covid-19 Pandemic on Device Usage and User
  Authentication}}. In \bibinfo{booktitle}{\emph{CHI Conference on Human
  Factors in Computing Systems Extended Abstracts}}. \bibinfo{pages}{1--7}.
\newblock


\bibitem[Saad et~al\mbox{.}(2023)]%
        {hotfoot}
\bibfield{author}{\bibinfo{person}{Alia Saad}, \bibinfo{person}{Kian Izadi},
  \bibinfo{person}{Anam Ahmad~Khan}, \bibinfo{person}{Pascal Knierim},
  \bibinfo{person}{Stefan Schneegass}, \bibinfo{person}{Florian Alt}, {and}
  \bibinfo{person}{Yomna Abdelrahman}.} \bibinfo{year}{2023}\natexlab{}.
\newblock \showarticletitle{HotFoot: Foot-Based User Identification Using
  Thermal Imaging}. In \bibinfo{booktitle}{\emph{Proceedings of the 2023 CHI
  Conference on Human Factors in Computing Systems}} (Hamburg, Germany)
  \emph{(\bibinfo{series}{CHI '23})}. \bibinfo{publisher}{Association for
  Computing Machinery}, \bibinfo{address}{New York, NY, USA}, Article
  \bibinfo{articleno}{262}, \bibinfo{numpages}{13}~pages.
\newblock
\showISBNx{9781450394215}
\urldef\tempurl%
\url{https://doi.org/10.1145/3544548.3580924}
\showDOI{\tempurl}


\bibitem[Samek et~al\mbox{.}(2017)]%
        {wojciech2017explainable}
\bibfield{author}{\bibinfo{person}{Wojciech Samek}, \bibinfo{person}{Thomas
  Wiegand}, {and} \bibinfo{person}{Klaus{-}Robert M{\"{u}}ller}.}
  \bibinfo{year}{2017}\natexlab{}.
\newblock \showarticletitle{Explainable Artificial Intelligence: Understanding,
  Visualizing and Interpreting Deep Learning Models}.
\newblock \bibinfo{journal}{\emph{CoRR}}  \bibinfo{volume}{abs/1708.08296}
  (\bibinfo{year}{2017}).
\newblock
\showeprint[arXiv]{1708.08296}
\urldef\tempurl%
\url{http://arxiv.org/abs/1708.08296}
\showURL{%
\tempurl}


\bibitem[Seitz and Hussmann(2017)]%
        {seitz2017pasdjo}
\bibfield{author}{\bibinfo{person}{Tobias Seitz} {and}
  \bibinfo{person}{Heinrich Hussmann}.} \bibinfo{year}{2017}\natexlab{}.
\newblock \showarticletitle{PASDJO: quantifying password strength perceptions
  with an online game}. In \bibinfo{booktitle}{\emph{Proceedings of the 29th
  Australian Conference on Computer-Human Interaction}}.
  \bibinfo{pages}{117--125}.
\newblock


\bibitem[Sepas-Moghaddam and Etemad(2022)]%
        {sepas2022deep}
\bibfield{author}{\bibinfo{person}{Alireza Sepas-Moghaddam} {and}
  \bibinfo{person}{Ali Etemad}.} \bibinfo{year}{2022}\natexlab{}.
\newblock \showarticletitle{Deep gait recognition: A survey}.
\newblock \bibinfo{journal}{\emph{IEEE transactions on pattern analysis and
  machine intelligence}} \bibinfo{volume}{45}, \bibinfo{number}{1}
  (\bibinfo{year}{2022}), \bibinfo{pages}{264--284}.
\newblock


\bibitem[Sharif et~al\mbox{.}(2019)]%
        {sharif2019overview}
\bibfield{author}{\bibinfo{person}{Muhammad Sharif}, \bibinfo{person}{Mudassar
  Raza}, \bibinfo{person}{Jamal~Hussain Shah}, \bibinfo{person}{Mussarat
  Yasmin}, {and} \bibinfo{person}{Steven~Lawrence Fernandes}.}
  \bibinfo{year}{2019}\natexlab{}.
\newblock \showarticletitle{An overview of biometrics methods}.
\newblock \bibinfo{journal}{\emph{Handbook of multimedia information security:
  techniques and applications}} (\bibinfo{year}{2019}),
  \bibinfo{pages}{15--35}.
\newblock


\bibitem[Sieger et~al\mbox{.}(2010)]%
        {sieger2010poster}
\bibfield{author}{\bibinfo{person}{Hanul Sieger}, \bibinfo{person}{Niklas
  Kirschnick}, {and} \bibinfo{person}{Sebastian M{\"o}ller}.}
  \bibinfo{year}{2010}\natexlab{}.
\newblock \showarticletitle{Poster: User preferences for biometric
  authentication methods and graded security on mobile phones}. In
  \bibinfo{booktitle}{\emph{Symposium on usability, privacy, and security
  (SOUPS)}}. Citeseer.
\newblock


\bibitem[Singh et~al\mbox{.}(2018)]%
        {singh2018vision}
\bibfield{author}{\bibinfo{person}{Jasvinder~Pal Singh},
  \bibinfo{person}{Sanjeev Jain}, \bibinfo{person}{Sakshi Arora}, {and}
  \bibinfo{person}{Uday~Pratap Singh}.} \bibinfo{year}{2018}\natexlab{}.
\newblock \showarticletitle{Vision-based gait recognition: A survey}.
\newblock \bibinfo{journal}{\emph{Ieee Access}}  \bibinfo{volume}{6}
  (\bibinfo{year}{2018}), \bibinfo{pages}{70497--70527}.
\newblock


\bibitem[Sprager and Juric(2015)]%
        {sprager2015inertial}
\bibfield{author}{\bibinfo{person}{Sebastijan Sprager} {and}
  \bibinfo{person}{Matjaz~B Juric}.} \bibinfo{year}{2015}\natexlab{}.
\newblock \showarticletitle{Inertial sensor-based gait recognition: A review}.
\newblock \bibinfo{journal}{\emph{Sensors}} \bibinfo{volume}{15},
  \bibinfo{number}{9} (\bibinfo{year}{2015}), \bibinfo{pages}{22089--22127}.
\newblock


\bibitem[Stylios et~al\mbox{.}(2022)]%
        {stylios2022biogames}
\bibfield{author}{\bibinfo{person}{Ioannis Stylios}, \bibinfo{person}{Spyros
  Kokolakis}, \bibinfo{person}{Andreas Skalkos}, {and}
  \bibinfo{person}{Sotirios Chatzis}.} \bibinfo{year}{2022}\natexlab{}.
\newblock \showarticletitle{BioGames: a new paradigm and a behavioral
  biometrics collection tool for research purposes}.
\newblock \bibinfo{journal}{\emph{Information \& Computer Security}}
  \bibinfo{volume}{30}, \bibinfo{number}{2} (\bibinfo{year}{2022}),
  \bibinfo{pages}{243--254}.
\newblock


\bibitem[Sundararajan and Woodard(2018)]%
        {sundararajan2018deep}
\bibfield{author}{\bibinfo{person}{Kalaivani Sundararajan} {and}
  \bibinfo{person}{Damon~L Woodard}.} \bibinfo{year}{2018}\natexlab{}.
\newblock \showarticletitle{Deep learning for biometrics: A survey}.
\newblock \bibinfo{journal}{\emph{ACM Computing Surveys (CSUR)}}
  \bibinfo{volume}{51}, \bibinfo{number}{3} (\bibinfo{year}{2018}),
  \bibinfo{pages}{1--34}.
\newblock


\bibitem[Taskiran et~al\mbox{.}(2020)]%
        {taskiran2020face}
\bibfield{author}{\bibinfo{person}{Murat Taskiran}, \bibinfo{person}{Nihan
  Kahraman}, {and} \bibinfo{person}{Cigdem~Eroglu Erdem}.}
  \bibinfo{year}{2020}\natexlab{}.
\newblock \showarticletitle{Face recognition: Past, present and future (a
  review)}.
\newblock \bibinfo{journal}{\emph{Digital Signal Processing}}
  \bibinfo{volume}{106} (\bibinfo{year}{2020}), \bibinfo{pages}{102809}.
\newblock


\bibitem[Teh et~al\mbox{.}(2013)]%
        {teh2013survey}
\bibfield{author}{\bibinfo{person}{Pin~Shen Teh}, \bibinfo{person}{Andrew
  Beng~Jin Teoh}, {and} \bibinfo{person}{Shigang Yue}.}
  \bibinfo{year}{2013}\natexlab{}.
\newblock \showarticletitle{A survey of keystroke dynamics biometrics}.
\newblock \bibinfo{journal}{\emph{The Scientific World Journal}}
  \bibinfo{volume}{2013} (\bibinfo{year}{2013}).
\newblock


\bibitem[Teh et~al\mbox{.}(2016)]%
        {teh2016survey}
\bibfield{author}{\bibinfo{person}{Pin~Shen Teh}, \bibinfo{person}{Ning Zhang},
  \bibinfo{person}{Andrew Beng~Jin Teoh}, {and} \bibinfo{person}{Ke Chen}.}
  \bibinfo{year}{2016}\natexlab{}.
\newblock \showarticletitle{A survey on touch dynamics authentication in mobile
  devices}.
\newblock \bibinfo{journal}{\emph{Computers \& Security}}  \bibinfo{volume}{59}
  (\bibinfo{year}{2016}), \bibinfo{pages}{210--235}.
\newblock


\bibitem[Tolba et~al\mbox{.}(2006)]%
        {tolba2006face}
\bibfield{author}{\bibinfo{person}{AS Tolba}, \bibinfo{person}{AH El-Baz},
  {and} \bibinfo{person}{AA El-Harby}.} \bibinfo{year}{2006}\natexlab{}.
\newblock \showarticletitle{Face recognition: A literature review}.
\newblock \bibinfo{journal}{\emph{International Journal of Signal Processing}}
  \bibinfo{volume}{2}, \bibinfo{number}{2} (\bibinfo{year}{2006}),
  \bibinfo{pages}{88--103}.
\newblock


\bibitem[Vazquez-Fernandez and Gonzalez-Jimenez(2016)]%
        {vazquez2016face}
\bibfield{author}{\bibinfo{person}{Esteban Vazquez-Fernandez} {and}
  \bibinfo{person}{Daniel Gonzalez-Jimenez}.} \bibinfo{year}{2016}\natexlab{}.
\newblock \showarticletitle{Face recognition for authentication on mobile
  devices}.
\newblock \bibinfo{journal}{\emph{Image and Vision Computing}}
  \bibinfo{volume}{55} (\bibinfo{year}{2016}), \bibinfo{pages}{31--33}.
\newblock
\showISSN{0262-8856}
\urldef\tempurl%
\url{https://doi.org/10.1016/j.imavis.2016.03.018}
\showDOI{\tempurl}


\bibitem[Vera-Rodriguez et~al\mbox{.}(2013)]%
        {vera2013gait}
\bibfield{author}{\bibinfo{person}{Ruben Vera-Rodriguez},
  \bibinfo{person}{John~S.D. Mason}, \bibinfo{person}{Julian Fierrez}, {and}
  \bibinfo{person}{Javier Ortega-Garcia}.} \bibinfo{year}{2013}\natexlab{}.
\newblock \showarticletitle{Comparative Analysis and Fusion of Spatiotemporal
  Information for Footstep Recognition}.
\newblock \bibinfo{journal}{\emph{IEEE Transactions on Pattern Analysis and
  Machine Intelligence}} \bibinfo{volume}{35}, \bibinfo{number}{4}
  (\bibinfo{year}{2013}), \bibinfo{pages}{823--834}.
\newblock
\urldef\tempurl%
\url{https://doi.org/10.1109/TPAMI.2012.164}
\showDOI{\tempurl}


\bibitem[Wan et~al\mbox{.}(2018)]%
        {wan2018survey}
\bibfield{author}{\bibinfo{person}{Changsheng Wan}, \bibinfo{person}{Li Wang},
  {and} \bibinfo{person}{Vir~V Phoha}.} \bibinfo{year}{2018}\natexlab{}.
\newblock \showarticletitle{A survey on gait recognition}.
\newblock \bibinfo{journal}{\emph{ACM Computing Surveys (CSUR)}}
  \bibinfo{volume}{51}, \bibinfo{number}{5} (\bibinfo{year}{2018}),
  \bibinfo{pages}{1--35}.
\newblock


\bibitem[Wang et~al\mbox{.}(2017)]%
        {wang2017single}
\bibfield{author}{\bibinfo{person}{Xinnian Wang}, \bibinfo{person}{Huiyu Wang},
  \bibinfo{person}{Qi Cheng}, \bibinfo{person}{Namusisi~Linda Nankabirwa},
  {and} \bibinfo{person}{Tao Zhang}.} \bibinfo{year}{2017}\natexlab{}.
\newblock \showarticletitle{Single 2D pressure footprint based person
  identification}. In \bibinfo{booktitle}{\emph{2017 IEEE International Joint
  Conference on Biometrics (IJCB)}}. IEEE, \bibinfo{pages}{413--419}.
\newblock


\bibitem[Wash(2010)]%
        {wash2010soups}
\bibfield{author}{\bibinfo{person}{Rick Wash}.}
  \bibinfo{year}{2010}\natexlab{}.
\newblock \showarticletitle{Folk Models of Home Computer Security}. In
  \bibinfo{booktitle}{\emph{Proceedings of the Sixth Symposium on Usable
  Privacy and Security}} (Redmond, Washington, USA)
  \emph{(\bibinfo{series}{SOUPS '10})}. \bibinfo{publisher}{Association for
  Computing Machinery}, \bibinfo{address}{New York, NY, USA}, Article
  \bibinfo{articleno}{11}, \bibinfo{numpages}{16}~pages.
\newblock
\showISBNx{9781450302647}
\urldef\tempurl%
\url{https://doi.org/10.1145/1837110.1837125}
\showDOI{\tempurl}


\bibitem[Yampolskiy and Govindaraju(2008)]%
        {yampolskiy2008behavioural}
\bibfield{author}{\bibinfo{person}{Roman~V Yampolskiy} {and}
  \bibinfo{person}{Venu Govindaraju}.} \bibinfo{year}{2008}\natexlab{}.
\newblock \showarticletitle{Behavioural biometrics: a survey and
  classification}.
\newblock \bibinfo{journal}{\emph{International Journal of Biometrics}}
  \bibinfo{volume}{1}, \bibinfo{number}{1} (\bibinfo{year}{2008}),
  \bibinfo{pages}{81--113}.
\newblock


\bibitem[Yang et~al\mbox{.}(2019)]%
        {yang2019security}
\bibfield{author}{\bibinfo{person}{Wencheng Yang}, \bibinfo{person}{Song Wang},
  \bibinfo{person}{Jiankun Hu}, \bibinfo{person}{Guanglou Zheng}, {and}
  \bibinfo{person}{Craig Valli}.} \bibinfo{year}{2019}\natexlab{}.
\newblock \showarticletitle{Security and accuracy of fingerprint-based
  biometrics: A review}.
\newblock \bibinfo{journal}{\emph{Symmetry}} \bibinfo{volume}{11},
  \bibinfo{number}{2} (\bibinfo{year}{2019}), \bibinfo{pages}{141}.
\newblock


\bibitem[Yapo and Weiss(2018)]%
        {yapo2018ethical}
\bibfield{author}{\bibinfo{person}{Adrienne Yapo} {and} \bibinfo{person}{Joseph
  Weiss}.} \bibinfo{year}{2018}\natexlab{}.
\newblock \showarticletitle{Ethical implications of bias in machine learning}.
  In \bibinfo{booktitle}{\emph{Proceedings of the 51st Hawaii International
  Conference on System Sciences}}.
\newblock


\bibitem[Zhao et~al\mbox{.}(2003)]%
        {zhao2003face}
\bibfield{author}{\bibinfo{person}{Wenyi Zhao}, \bibinfo{person}{Rama
  Chellappa}, \bibinfo{person}{P~Jonathon Phillips}, {and}
  \bibinfo{person}{Azriel Rosenfeld}.} \bibinfo{year}{2003}\natexlab{}.
\newblock \showarticletitle{Face recognition: A literature survey}.
\newblock \bibinfo{journal}{\emph{ACM computing surveys (CSUR)}}
  \bibinfo{volume}{35}, \bibinfo{number}{4} (\bibinfo{year}{2003}),
  \bibinfo{pages}{399--458}.
\newblock


\bibitem[Zhong et~al\mbox{.}(2019)]%
        {zhong2019decade}
\bibfield{author}{\bibinfo{person}{Dexing Zhong}, \bibinfo{person}{Xuefeng Du},
  {and} \bibinfo{person}{Kuncai Zhong}.} \bibinfo{year}{2019}\natexlab{}.
\newblock \showarticletitle{Decade progress of palmprint recognition: A brief
  survey}.
\newblock \bibinfo{journal}{\emph{Neurocomputing}}  \bibinfo{volume}{328}
  (\bibinfo{year}{2019}), \bibinfo{pages}{16--28}.
\newblock


\end{thebibliography}

\appendix
\newpage
\section{Questionnaire}\label{app:questionnaire}

In this section, we list the full set of questions asked in our survey. Wherever an \_ appears it was replaced by the type of biometric (physiological/behavioral) in the corresponding part of the survey.

\subsection{Questions in Part \textit{B} (Demographics)}

Please answer the following short questions about yourself.

\begin{enumerate}
    \item How old are you?
    \item What is your gender?
    \item What is your profession?
    \item How would you rate your technical knowledge? ( 1 (no knowledge) to 5 (very proficient))
\end{enumerate}

\subsection{Questions in Part \textit{C} (Biometric Methods)}

In the following, you will see several questions about biometric methods.

Are you familiar with the concept of (physiological) biometric methods? 
\begin{enumerate}
    \item[\textit{Yes}] Please give a short definition/explanation. Please give a short explanation of how it works
    \item[\textit{No}] Please think about what it could be and answer with your thoughts
\end{enumerate}

\subsection{Questions in Part D (Definition)}

With this definition [see Section \ref{sec:survey_structure}] in mind, please answer the following questions:

\begin{enumerate}
    \item Please name all physiological biometric methods that you know/have heard of.
    \item Please name all behavioral biometric methods that you know/have heard of.
    \item Do you use biometric methods in your everyday life?
    \begin{enumerate}
        \item[\textit{Yes}] Please indicate all of them; why do you use them.
        \item[\textit{No}] why not?
    \end{enumerate}
    \item Beyond the methods that you know about, which physiological characteristics 
	could you imagine to be used for biometrics? Also, give an application in case you 
	have one in mind. 
    \item Beyond the methods that you know about, which behavioral characteristics 
	could you imagine to be used for biometrics? Also, give an application in case you 
	have one in mind. 
\end{enumerate}

\subsection{Questions in Part \textit{E} (Authentication)}

Please answer the following questions about your authentication behavior:

\begin{enumerate}
    \item Do you own and regularly use a smartphone?
    \item What operating system does your smartphone use (e.g. Android, IOS, Windows).
    \item What (primary) authentication scheme do you use (i.e. how do you usually unlock 
	your device)? Please select only one option to indicate your primary authentication scheme.
 \item In case you use a biometric method, what is your fallback authentication scheme? 
	Fallback refers to the method that you have to enter in case your primary 
	authentication method fails or in some cases on a regular basis (commonly 3 days)
\end{enumerate}

\subsection{Likert Statements in Part \textit{F} (Biometric Perception)}\label{app:likerts}

Please rate the following statements about physiological biometric methods. There are no right or wrong answers. This is  only about your perception. 

\begin{enumerate}
    \item Compared to a pin/password, using \_ biometrics makes authentication faster.
    \item  \_ biometrics methods are reliable.
    \item  \_ biometric methods are easy to use.
    \item  Performance of \_ biometric methods (security, errors,...) is equal for all users.
    \item  Compared to a pin/password, using \_ biometrics makes authentication more secure.
    \item  \_ biometrics are well suited to protect my personal data.
    \item  \_ biometrics can be faked.
    \item  I have concerns about my privacy when using \_ biometrics.
    \item  I am concerned that someone might hack my device/account when using \_ biometrics.
    \item  I am concerned that I might have no access to my device/account when using \_ biometrics.
\end{enumerate}

\subsection{Questions in Part \textit{G} (Performance \& Security)} 

Please answer the following questions about your perception of \_ biometrics.

\begin{enumerate}
    \item Do you think someone else could access your device/account if you protect it with \_ biometrics?
    \item Do you think changes in your \_ have an impact on \_ biometric systems?
    \item  What do/would you do, in case your \_ biometric system does not work  (i.e. you are unable to authenticate with it)?
    \item What do you think would happen if someone hacked your \_ biometric system (i.e. what would be the practical consequences)? What would you do?
    \item If you were to attack a \_ biometric system, how would you do it?
    \item What do you think can be done to make \_ biometric systems more secure? 
\end{enumerate}

\subsection{Questions in Part \textit{H} (Conclusion)} 

Thank you for your participation. Please answer these final questions and enter your email address below in case you want to participate in the raffle. 

\begin{enumerate}
    \item Are there any questions that you would have liked to answer differently after 
	completing the questionnaire (but were not able to do so because going back was 
	not allowed)? If so: which ones and what changed?
    \item Are there any other remarks that you would like to make?
    \item Please indicate your email address in case you want to participate in the raffle. Your 
	address will not be associated with your answers and will be deleted after the 
	winners were determined.
\end{enumerate}

Thank you for your support.

\pagebreak
\section{Codebook}\label{app:codebook}

Themes and groups as they resulted from our thematic analysis. For each group, we give counts of unique participants mentioning it. Counts are divided by the iteration of the survey (1: 2019, 2:2023).

\begin{table*}[h]
    \centering
\begin{tabular}{llrrr}
\toprule
    & \textbf{Round} &   1 &   2 &  All \\
\textbf{Theme} & Group &     &     &      \\

\midrule
\textbf{Definition}   
    & Features Mentioned  &  45 &  19 &   64 \\
    & Correct or Related Action Mentioned &  32 &  15 &   47 \\
    & Definition by Example &  24 &  12 &   36 \\     
    & Missing Knowledge &  48 &  34 &   82 \\
    & No Answer & 53 & 42 & 95\\

\midrule
\textbf{Usage of Biometrics} 
    & Examples of Usage of Biometric Methods &  29 &  27 &   56 \\
     & Devices/Usecases for which Biometric Methods &   19 &  23 &   42 \\ 
    & Reasons for using Biometric Methods &   4 &   7 &    11 \\    
    & Examples of Avoidance of Biometric Methods &   1 &   3 &    4 \\
    & Reasons for not using Biometric Methods &  14 &   5 &   19 \\

\midrule
\textbf{Attacks \& Challenges} 
    & Non-Malicious: Generic &  18 &  23 &   41 \\
    & Non-Malicious: Physical &  20 &   24 &   44 \\
    & Non-Malicious: Behavioural &   2 &   7 &    9 \\ 
    & Attack: Software based &  13 &  10 &   23 \\
    & Attack: Force &   7 &   9 &   16 \\
    & Attack: Immitation/Replay &  36 &  36 &   72 \\
    & Attack: Other &  11 &   5 &   16 \\
    & Perceived Weakness &  29 &  20 &   49 \\
    & No Impact &   9 &   4 &   13 \\    
    
\midrule    
\textbf{Consequences} 
    & Action: Fall back to other Method &  50 &  43 &   93 \\
    & Action: Damage Control &  24 &  21 &   45 \\
    & Action: Seek Support &  24 &  10 &   34 \\
    & Action: Other &   5 &   3 &    8 \\ 
    & Damage: Access and Missuse &  17 &  18 &   35 \\
    & Damage: Loss &  17 &  11 &   28 \\   
    & Damage: None &   5 &   6 &   11 \\      
    & Other Consequence &  12 &   3 &   15 \\

\midrule
\textbf{Future Suggestions} 
    & Novel Biometrics &  37 &  34 &   71 \\   
\textbf{ \& Improvements}   
    & Future Applications &   7 &   5 &   12 \\ 
    & Improvements &  27 &  30 &   57 \\ 
    
\bottomrule
\end{tabular}
\end{table*}

\end{document}